\documentclass[usenatbib,twocolumn]{mnras} 
\usepackage[T1]{fontenc}
\usepackage{ae,aecompl}
\usepackage{graphicx}
\usepackage{hyperref}
\usepackage{amsmath}
\usepackage{amssymb}
\usepackage{mathtools}
\usepackage{bm}
\usepackage{cleveref}
\usepackage{widetext}
\usepackage{stfloats}

\usepackage{subcaption, caption}

\usepackage{float}
\usepackage{ dsfont }
\usepackage{ upgreek }

\usepackage{color,xcolor}

\newcommand{\W}{\mathcal{W}}
\DeclareMathOperator\arctanh{arctanh}

\title[Transport in the nuclear pasta]{ 
Anisotropic electron transport in the nuclear pasta phase }

\author[Pelicer et al.]{
    M.~R.~Pelicer$^1$
    \thanks{E-mail: m.reinke.pelicer@posgrad.ufsc.br},
    M.~Antonelli$^2$ 
    \thanks{E-mail: antonelli@lpccaen.in2p3.fr}, 
    D.~P.~Menezes$^1$ 
    \thanks{E-mail: debora.p.m.26@gmail.com},
    F.~Gulminelli$^2$
    \thanks{E-mail: gulminelli@lpccaen.in2p3.fr}
    \\ 
    \\
    $^1$Depto de F\'{\i}sica - CFM - Universidade Federal de Santa Catarina  Florian\'opolis - SC - CP. 476 - CEP 88.040 - 900 - Brazil
    \\
    $^2$CNRS and ENSICAEN, Laboratoire de Physique Corpusculaire, 14050 Caen, France
    }

\begin{document}
    %
    %
    \pagerange{\pageref{firstpage}--\pageref{lastpage}} \pubyear{2022}
    \maketitle
    \label{firstpage}

\begin{abstract}
The presence of nuclear pasta is expected to modify the transport properties in the mantle of neutron stars. The non-spherical geometry of the pasta nuclear clusters leads to anisotropies in the collision frequencies, impacting the thermal and electrical conductivity.  
We derive analytical expressions for the anisotropic collision frequencies using the Boltzmann equation in the relaxation time approximation.
The  average parallel, perpendicular and Hall electrical conductivities are computed in the high-temperature regime above crustal melting, considering incoherent elastic electron-pasta scattering and randomly oriented pasta structures. Numerical values are obtained at different densities and temperatures by using the IUFSU parametrization of the non-linear Walecka model to determine the crustal structure. 
We find that the anisotropy of the collision frequencies grows with the length of the pasta structures and, independently of the magnetic field, the presence of rod and slab phases decreases the conductivity by more than one order of magnitude.
Our numerical results indicate that, even if the pasta structures might survive above the crustal melting point, no strong anisotropies are to be expected in the conduction properties in this temperature regime, even in the presence of a very high magnetic field.
\end{abstract}



\section{Introduction}

Observations related to the thermal, magnetic and spin evolution of neutron stars can provide us with indirect information on the transport properties of ultra-dense matter, e.g.~\citet{Horowitz_prl2015,montoli_bayes,Potekhin2021A&A}. 
In principle, the observations must be compared with simulations by properly modelling the coupled magneto-thermal evolution. Hence, models are necessary for the microscopic processes that give rise to the thermal and electric conductivities and viscosity throughout the star \citep{Page_Reddy_2012arXiv1201,chamel_2008LRR,Schmitt2018}, which are then used as inputs to the macroscopic simulations, see~\citet{2018MNRAS.473.2771B,Pons_2019,camelio2022arXiv}.

In the crust, transport properties are determined by the scattering of electrons by other electrons, ionic impurities and phonons in the crystal lattice. 
Electron-ion scattering dominates at the lowest densities and has been extensively studied~\citep{flowers1976, 1980SvA_yakovlev, 1984MNRAS.209..511N, Baiko:1998xk,Potekhin:1999yv, Chugunov:2005nc,PhysRevLett.102.091101}.
In the inner crust at temperatures $T<10^7\,$K, thermal conductivity due to degenerate electron-electron Coulomb scattering dominates over the contribution due to electron-phonon scattering \citep{Shternin_PhysRevD_06} and becomes competitive with the electron conductivity due to the scattering of electrons by impurity ions \citep{chamel_2008LRR}. 

The situation gets more complicated in the innermost part of the crust, where it might be energetically favourable for the ions composing the crystal lattice to deform in complex structures known as ``pasta''~\citep{PhysRevLett.50.2066, 10.1143/PTP.71.320, Oyamatsu:1993zz}.  
Classical molecular dynamics simulations suggest that this matter is disordered and amorphous and that different shapes might coexist at a given depth of the star, due to the small energy barriers between them~\citep{PhysRevC.90.055805,Horowitz_prl2015,2021PhRvC.103e5810C,newton_glassy_pasta}. This shape coexistence has been validated by relativistic mean field (RMF) calculations~\citep{PhysRevC.104.L022801}. 
In the case of a disordered and amorphous inner crust with randomly distributed nuclear clusters of different sizes~\citep{Carreau2020,Potekhin2021A&A} and geometries~\citep{PhysRevC.104.L022801}, the main mechanism of charge and heat transport is given by uncorrelated scattering processes between the electrons and the clusters, which play a role similar to one of the lattice impurities in a crystallized phase.

Regarding the possible astrophysical consequences, a high impurity parameter in the inner crust raises the electrical resistivity of the star, decreasing steeply the magnetic field after a certain age and thus the spin-down. This may explain the very small number of isolated X-ray pulsars with spin periods larger than 12s~\citep{Pons2013,newton2013taste,Hambarayan2017,Tan2018}.  
The high impurity also lowers the thermal conductivity, leading to a better fit of the late-time cooling of the binary MXB 1659-29~\citep{Horowitz_prl2015,Deibel2017}. 
Furthermore, the presence of the pasta layers modifies the so-called mutual friction force between the nuclear clusters and the neutron superfluid, with consequences on the pulsar glitch phenomenon \citep{2020MNRAS499}. Gravitational waves~\citep{Horowitz:2009vm}, quasi-periodic oscillations~\citep{Sotani:2012utk}, quasi-persistent sources of SXRTs and giant flares due to the relaxation of the crust after heat deposition and neutrino emissivity~\citep{PhysRevC.83.035803, PhysRevC.70.065806, Lin:2020nxy} are also influenced by the presence of an amorphous layer in the inner crust. 

In the presence of a strong $B$ field, electron transport is anisotropic, as the field bends the electron trajectories in the orthogonal plane and suppresses electron transport across the  direction of $\bm{B}$, e.g. \citep{chamel_2008LRR}. 
This argument considers that the only source of anisotropy is given by the $\bm{B}$ direction. However, the spherical symmetry of nuclear clusters is spontaneously broken in the pasta layers, leading to additional anisotropies already at the level of the microscopic scattering process: in particular, \citet{Yakovlev2016} has shown that, even in the case of random orientation of the pasta structures, anisotropic scattering can modify the transport properties. 

In the analysis of \citet{Yakovlev2016}, the scattering rates along and across the pasta symmetry axis were taken as free parameters. While molecular dynamics has been able to provide estimates of the transport properties in the inner crust, by taking the angular average of the effective structure factor of the charge distribution~\citep{Horowitz:2008vf, Horowitz_prl2015, Nandi_2018}, to our knowledge no estimation of the different collision frequencies that arise due to the pasta anisotropic shapes has been performed to date.

The existing microscopic simulations of the finite temperature pasta ~\citep{PhysRevC.90.055805,Horowitz_prl2015,2021PhRvC.103e5810C,newton_glassy_pasta,Nandi_2018} are typically done at fixed proton fraction and high temperatures $T\ge 10^{10}$ K, thermodynamic conditions that are especially aimed at the description of proto-neutron stars formed in supernova events. 
In these conditions, it appears from those calculations that the distribution of baryonic matter is strongly disordered, and one might expect that anisotropies should not have a strong effect on the transport properties.  
On the other hand, in the case of neutron star binaries and soft X-ray transients, the inner crust is close to $\beta$-equilibrium and temperatures are one or two orders of magnitude lower, which might preserve both the peculiar pasta geometrical shapes and the lattice quasi-long range order, potentially leading to a strong anisotropy of the scattering rates, as assumed by ~\citet{Yakovlev2016}.
 
In this paper, we show how the anisotropic collision frequencies can be calculated from the Boltzmann equation in the relaxation time approximation, in the case of elastic scattering of ultra-relativistic degenerate electrons off pasta structures. We limit ourselves to the hypothesis of incoherent scattering sources following the Matthiessen rule \citep{Schmitt2018,Heiselberg_PhysRevD_93,Shternin_PhysRevD_06}. 
Based on the behaviour of the static structure factor,  we argue that this hypothesis should be valid in the high-temperature regime above crustal melting.  

The paper is organized as follows.
In Sec.~\ref{sec:nuanisotropic} we calculate the general anisotropic collision frequencies.
The collision integral and the transition matrix elements are first expanded in the spherical harmonics basis in Sec.~\ref{sec:expansion}. 
Then, to extract the physical real collision frequencies, in Sec.~\ref{sec:nul1} we consider the lowest order (dipole) deviation from equilibrium and take advantage of the axial symmetry of the pasta phase.
The contribution of the collision integral to the conductivity is given in terms of axial and perpendicular collision frequencies, in agreement with the analysis of~\citet{Yakovlev2016}. 
Analytical expressions for the conductivity matrix are given in Sec.~\ref{sec:nuap} for the case of a liquid, disordered, pasta phase. In Sec.~\ref{sec:incoherent}, the transition matrix is numerically evaluated in the temperature domain of validity of our approximations. 
The conductivity tensor with and without magnetic field is finally obtained in Sec.~\ref{sec:tensor}. 
To illustrate the formalism and give quantitative estimations of the transport coefficients, in Sec.~\ref{sec:pasta} we present numerical calculations for the collision frequencies and conductivity for different densities and $B$ values in the high-temperature regime.
Conclusions are presented in Sec.~\ref{sec:conclusions}.

All the numerical estimates reported in this paper are obtained using the IUFSU parametrization of the RMF approach for the crustal composition, see~\citealt{IUFSU, PRC85}, but our expressions can be employed with any nuclear physics model that gives the static structure of the crust. In particular, while our quantitative numerical results might be model dependent, the qualitative conclusions remain valid for any other realistic equation of state model. 

We use natural units $\hbar = c = k_B= 1$ all over the paper. 

\section{Relaxation time approximation for anisotropic elastic collisions} 
\label{sec:nuanisotropic}

The thermal and electrical electron conductivities due to electron-ion scattering have been calculated in a wide range of temperatures $T$ and electron densities $n_e$, see e.g. \citet{Potekhin_1999A&A}: for homogeneous media, and in the absence of a magnetic field, they are expressed in terms of the effective collision frequencies $\nu_{\sigma,\kappa}$ as 
\begin{equation}
    \label{eq:sigmakappa_iso}
    {\sigma} = \frac{e^2 n_e}{ m_e^\ast \nu_\sigma} 
    \qquad \qquad
    {\kappa} = \frac{\pi^2 T n_e}{3 m_e^\ast \nu_\kappa} \, ,
\end{equation}
where $m_e^\ast$ is the effective electron mass, and in the liquid regime $\nu_\sigma=\nu_\kappa\equiv\nu$, with the collision frequency $\nu$ defined as the inverse of the relaxation time, $\nu = 1/\tau$.
Because of the isotropy assumption, the collision frequencies only depend on the modulus of the momentum transfer according to the general expression~\citep{flowers1976, 1980SvA_yakovlev, 1984MNRAS.209..511N}:
\begin{equation}
    \label{eq:nu_iso}
    \nu = \frac{4 \pi n_i e^4 Z^2}{v_F p_F^2} \int_{0}^{2 p_F} \frac{d q}{q}  \left( 1- \frac{q^2}{4 \upepsilon_{F}^2}\right)  \frac{F^2(q)}{\varepsilon^2(q)} S(q) \,  ,
\end{equation}
where $F(q)$ is the ion form factor, $\epsilon(q)$ is the dielectric function, $S(q)$ is the effective structure factor that accounts for ion correlations, and $v_F$, $p_F$, and $\upepsilon_{F}$ are the Fermi velocity, momentum and energy respectively. 
Unfortunately, eqs~\eqref{eq:sigmakappa_iso} and \eqref{eq:nu_iso} cannot be straightforwardly generalized to the case of anisotropic scatterings.  
To derive the anisotropic collision frequencies, we consider a multipole expansion of the Boltzmann equation in the relaxation time approximation, as we detail below. 

\subsection{Anisotropic case: expansion in spherical harmonics}\label{sec:expansion}
 
We consider a strongly degenerate {relativistic} electron gas with position-dependent temperature and chemical potential fields $T(\bm{ r})$ and $\mu(\bm{ r})$ in a constant external magnetic field $\bm{ B}$ and a weak electric field $\bm{ E}$.  
Assuming that the gas is only slightly out of equilibrium, we can write its distribution function as $f(\bm{r}, \bm{p}, t)= f_0(\bm{r}, \upepsilon_p) + \delta f(\upepsilon_p)$, where $\bm{r}$,  $\bm{v}$ and  $\bm{p}$ are the electron position, velocity and momentum, respectively, with the latter given by $\bm{p} = \upepsilon_p \bm{v}$. 
The Fermi-Dirac function  $f_0$ is given by 
\begin{equation}
    f_0(\bm{r}, \upepsilon_p) = \left[ 1 + \exp\left(\frac{\upepsilon_p - \mu(\bm{r})}{T(\bm{r})} \right) \right]^{-1}.
\end{equation}
The deviation from equilibrium can be found by solving the linearized Boltzmann equation \citep{Heiselberg_PhysRevD_93,Shternin_PhysRevD_06}
\begin{equation}\label{eq:boltzmann_linear}
\begin{split}
      \left( -\frac{\partial f_0}{\partial \upepsilon_p} \right) \bm{v} \cdot \Bigg[ \nabla \mu + e  \bm{E} & + \frac{\upepsilon_p - \mu}{ T} \nabla T \Bigg] \\
      &-e (\bm{v} \times \bm{B}) \cdot \frac{\partial \delta f}{\partial \bm{p}}  = I[f] \, ,
\end{split}
\end{equation}
where $I[f]$ is the collision integral that can be written as
\begin{equation}\label{eq:I}
\begin{split}
    I[f] = \int \frac{d^3 \bm{p}\, '}{(2 \pi)^3} &  \big[ \Gamma_{p'\rightarrow p}  f(\bm{p}\,') \left( 1- f (\bm{p} ) \right)  \\
    &- \Gamma_{p\rightarrow p'}  f(\bm{p}) \left( 1- f (\bm{p}\,' ) \right) \big] \, .
\end{split}
\end{equation}
Here, $\Gamma_{p\rightarrow p'}$ is the transition rate from an initial momentum  $\bm{p}$ to a final momentum ${\bm{p}}\,'$, {introduced} to account for electron scattering with any generic potential, and we have omitted the position dependencies as they do not affect the calculation. We restrict ourselves to elastic scatterings with a localized source for the potential, such that the following simplification applies:
\begin{equation}\label{eq:elastic_assumption}
    \Gamma_{p \rightarrow p'} = \Gamma_{p' \rightarrow p} = 2 \pi \delta(\upepsilon_p - \upepsilon_{p'}) W_{pp'},
\end{equation}
where $W_{pp'}$ is the transition matrix element, which we will write explicitly for the case of electron-pasta scattering in the next section. 
Taking into account that deviations from equilibrium are small, we can rewrite the collision integral, eq.~\eqref{eq:I}, as 
\begin{equation}\label{eq:I_el}
    I[f] = - 2 \pi \int \frac{d^3 \bm{p}\, '}{(2 \pi)^3}   
    \delta( \upepsilon_p - \upepsilon_{p'}) W_{p p'} 
    \left[ \delta f (\bm{p}) - \delta f (\bm{p}\,') \right],
\end{equation}
where the Fermi-Dirac terms coming from the different momenta have cancelled out due to the elasticity assumption in eq.~\eqref{eq:elastic_assumption}.

In isotropic scatterings, $W_{pp'}$ is a function of $q= |\bm{p} - {\bm{p}}'|$ and of the electron energy only. Since {in this work} we are dealing with general anisotropic scatterings, we will assume {the matrix elements} to be functions of the  solid angles of both incoming and outgoing electron momenta ($\bm{p}$ and ${\bm{p}}'$), as well as the energy $\upepsilon_p$, so there is no assumption of symmetry for the source of potential. 
The transition matrix can be expanded in the basis of spherical harmonics as
\begin{equation}
\label{eq:Wexpansion}
    W_{p \,p'} \left( \Omega_p, \Omega_{p'}, \upepsilon_p \right) =  \sum_{lm \, l'm'} \W_{lm \,  l'm'}(\upepsilon_p) Y_l^m (\Omega_p) Y_{l'}^{m'}(\Omega_{p'}),
\end{equation}
whereas the assumption of elasticity implies that $\Omega_p$ and $\Omega_{p'}$ are interchangeable, such that
\begin{equation}\label{eq:W_el}
    W_{pp'} = W_{p'p} \;\; \Longrightarrow \;\;  \W_{lm\,l'm'} = \W_{l'm'\, lm}.
\end{equation}

This is a generalization of the Legendre expansion used for scattering with isotropic targets -- see Sec. 3 in \citet{pines2018theory}. In App.~\ref{app:iso_limit} we show how the isotropic limit can be recovered from our calculation.

\begin{figure*}
    \centering\includegraphics[scale=0.8]{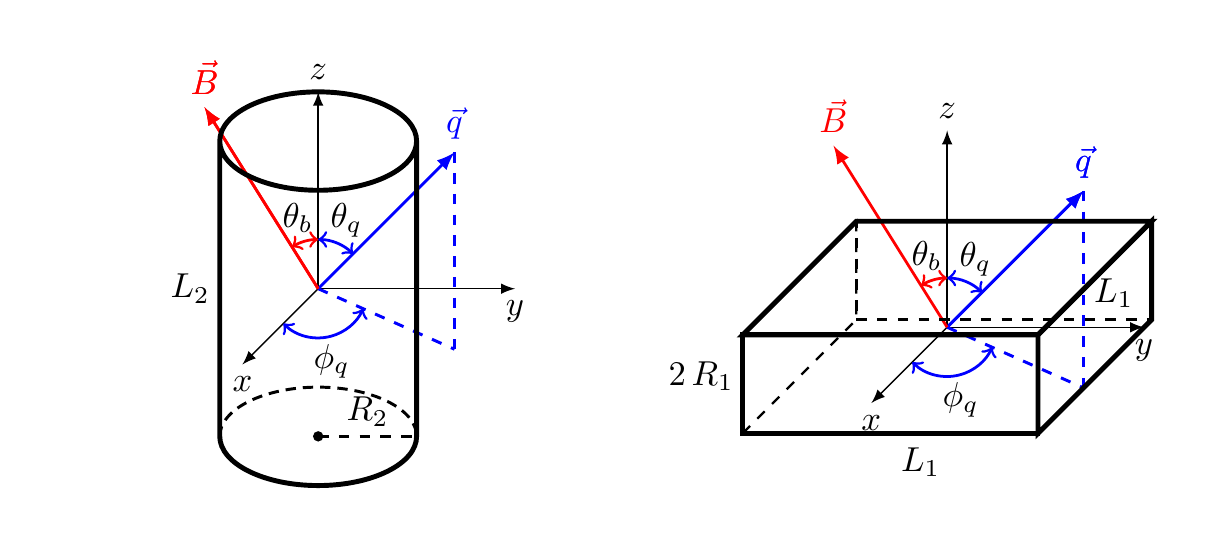}
    \caption{Cylindrical (rod and tube) and planar (slab) geometries of the nuclear pasta with $z$ as the symmetry axis.  The transferred momentum  vector $\bm{q}$ is drawn arbitrarily and the magnetic field $B$ lies in the $xz$--plane.}
    \label{fig:pasta}
\end{figure*}

 The deviation from equilibrium of the electron distribution is expanded as:
\begin{equation}\label{eq:df_sh}
    \delta f (\bm{p})= \sum_{lm} \delta f_{lm} (\upepsilon_p) Y_l^m (\Omega_p).
\end{equation}
Substitution of  eqs~\eqref{eq:Wexpansion} and \eqref{eq:df_sh} into eq.~\eqref{eq:I_el} allows us to use the orthogonality of spherical harmonics and the contraction rule 
\begin{flalign}
    Y_l^m(\Omega) & Y_{l'}^{m'}(\Omega)  = \sum_{L\, M} (-1)^M 
   \sqrt{\frac{(2L+1)(2l+1)(2l'+1)}{4\pi}} \nonumber \\
   & \times \begin{pmatrix}
l & l' & L\\
0 & 0 & 0
\end{pmatrix}
 \begin{pmatrix}
l & l' & L\\
m & m' & -M
\end{pmatrix} Y_L^M(\Omega)
\end{flalign}
to rewrite {the collision integral} as
\begin{widetext}
\begin{equation}\label{eq:I_nu}
\begin{split}
    I[f]  = -  \frac{p^2 }{4 \pi^2 v } \sum_{lm, l'm'} \delta f_{lm} 
    \bigg[ \W_{l'm'\;00}  & \sum_{LM} (-1)^M \sqrt{(2 l +1)(2l'+1)(2L+1)} \\
&\times\begin{pmatrix}
l & l' & L\\
0 & 0 & 0
\end{pmatrix}
  \begin{pmatrix}
l & l' & L\\
m & m' & -M
\end{pmatrix}
Y_L^M (\Omega_p)  - (-1)^m \W_{l'm'\,l-m} Y_{l'}^{m'}(\Omega_p) 
\bigg].
\end{split}
\end{equation}
\end{widetext}

\newpage
The 3-j Wigner symbols  
{\footnotesize $\begin{pmatrix}
l_1 & l_2 & l_3\\
m_1 & m_2 & m_3
\end{pmatrix} $} 
are invariant under even permutations of the columns and non-zero only if $m_1+m_2+m_3=0$, $|l_1-l_2|< l_3<l_1+l_2$ and $l_1+l_2+l_3$ is an integer~\citep{brink1968theory, edmonds2016angular}.

We define the anisotropic collision frequencies by expanding the collision integral linearly in $\delta f$, 
\begin{equation}\label{eq:I_sh}
    I[f ] = - \sum_{lm, l'm'} \delta f_{lm}\left(\upepsilon_p\right) \, \left[{\nu}\left(\upepsilon_p\right)\right]_{lm}^{l'm'}  Y_{l'}^{m'} (\Omega_p),
\end{equation}
Integration of eqs~\eqref{eq:I_nu} and~\eqref{eq:I_sh} in $\Omega_p$ yields
\begin{widetext}
\begin{equation}\label{eq:nu_final}
\left[\nu\right]_{lm}^{l'm'} =     \frac{p^2 }{4 \pi^2 v } \Bigg[(-1)^{m'} \sqrt{(2 l +1)(2l'+1)}  \sum_{LM} \W_{LM \;00} \sqrt{2 L+1}
\begin{pmatrix}
l & l' & L\\
0 & 0 & 0
\end{pmatrix} \begin{pmatrix}
L & l & l'\\
M & m & -m'
\end{pmatrix}
-  \; (-1)^m \, \W_{l'm'\,l-m} \Bigg].
\end{equation}
\end{widetext}
We can obtain a more compact form of this expression by using the Wigner–Eckart theorem and the spherical harmonics representation of irreducible tensor operators of rank~$l$~\citep{PhysRev.61.186, PhysRev.62.438},
\begin{equation}
    C^l_m = \sqrt{\frac{4 \pi}{2l+1}} Y_{lm} (\Omega) \, ,
\end{equation}
such that eq.~\eqref{eq:nu_final} becomes
\begin{equation}
    \label{eq:nu_final_irr}
    \begin{split}
    \left[\nu\right]_{lm}^{l'm'} = \frac{p^2 }{4 \pi^2 v } \Bigg[\sum_{LM} \W_{LM \;00}
    &  \sqrt{2 L+1} \langle l' m' | C_M^L | l m \rangle   \\
    & -  \; (-1)^m \, \W_{l'm'\,l-m} \Bigg] \,.
    \end{split}    
\end{equation}

\subsection{Derivation of the collision frequencies}
\label{sec:nul1}

To evaluate the collision frequencies in the pasta phase, we consider idealized rod and slab-like geometries, as expected in the basic liquid-drop modelling of the inner crust~\citep{PhysRevLett.50.2066, 10.1143/PTP.71.320}. 
These geometries and the definitions entering the calculations are sketched in Fig.~\ref{fig:pasta}. 

Equation \eqref{eq:nu_final_irr} is not yet a multipole expansion of the collision rates because the different expansion coefficients of the collision integral $\left[\nu\right]_{lm}^{l'm'}$ are complex numbers. This is due to the fact that both the electron distribution function and the collision integral are written on the basis of complex spherical harmonics. To relate eq.~\eqref{eq:nu_final_irr} to the physical quantities, we must rewrite eqs~\eqref{eq:df_sh} and \eqref{eq:I_sh} in terms of real coefficients. 

To do so, we notice that the coefficients $\W_{l'm'\,l m}$ in eq.~\eqref{eq:Wexpansion} are constrained by the symmetries of rods and slabs. Both geometries are invariant under inversion of the $z$-axis ($z \rightarrow -z$), implying that the only non-zero  $\W_{l'm'\,l m}$ are the ones having the sum $l+l'$ that is even. The sum  $m+ m'$ is constrained by the $xy$--plane symmetries: cylinders are invariant under arbitrary rotations: the non-zero $\W_{l'm'\,l m}$ are only those with $m+m'=0$; slabs are invariant over $\pi/2$ rotations, so the sum $m+m'$ must be a multiple of 4. 
To summarize, the $\W_{lm\, l'm'}$ are not zero if and only if:
\begin{equation}\label{eq:W_sym}
    \text{Rods} \, \left.
\begin{cases}
l+l'= 2 k\\
m+m'=0
\end{cases}
\right.
\qquad
    \text{Slabs}\,\left.
\begin{cases}
l+l'= 2 k \\
m+m'=4 k'\\
\end{cases}
\right.
\end{equation}
with  $k, k' \in \mathds{Z}$.
To further progress, we restrict ourselves to the case of electric and thermal conductivities.
Accounting for spin degeneracy, the electric current and heat flow are given by
\begin{equation}\label{eq:currents}
    \bm{j} = -2e \int \frac{d^3 \bm{p} }{(2 \pi)^3} \bm{v} \delta f \qquad \bm{q} = 2 \int \frac{d^3 \bm{p} }{(2 \pi)^3} \bm{v}  (\upepsilon - \mu) \delta f \, ,
\end{equation}
so that only the odd $l$ terms in the expansion \eqref{eq:df_sh} contribute to the integrals in eq.~\eqref{eq:currents}.
Moreover, in the relaxation time approximation, the collision integral, eq.~\eqref{eq:I_el}, is linear in $\delta f$. The left-hand side of eq.~\eqref{eq:boltzmann_linear} is linear in $\bm{p}$, implying that only the coefficient $l'=1$ in the expansion of the collision integral, eq.~\eqref{eq:I_sh} contributes to the currents\footnote{
    The multiplicity $l$ of the spherical harmonics coincides with the power of $\bm{p}$ in an equivalent expansion in homogeneous harmonic polynomials since they are isomorphic~\citep{gallier2013,freire2022}.
}. 
This is also discussed in depth in the case of isotropic scattering in~\citet{SYKES19701}, and mentioned in the case of pasta in~\citet{Schmitt2018}.

In the isotropic case, there is no mixing between the different terms of $I[f]$ in eq.~\eqref{eq:I_sh} and those of $\delta f$ in eq.~\eqref{eq:df_sh}. However, in the anisotropic case, the collision frequencies can mix the $l'=1$ contributions in eq.~\eqref{eq:I_sh} with the  $l=2, 3...$ components of $\delta f_{lm}$ in eq.~\eqref{eq:I_sh}.
For simplicity, we neglect such mixing and restrict ourselves to the most important contribution (see also \citealt{Schmitt2018}) by writing $\nu_{lm}^{l'm'} = \nu_{1m}^{1 m'} \delta_{l1} \delta_{l'1}$ in eq.~\eqref{eq:I_sh}. 
This approximate approach is probably good in the case of pasta, due to the symmetry rules in~\eqref{eq:W_sym}. 

We will show that the axial symmetry of the problem limits the number of physical collision frequencies to two: an axial frequency $\nu_a$, and a perpendicular one $\nu_p$.  
To do so, we need to rewrite the expansions eqs~\eqref{eq:df_sh} and~\eqref{eq:I_sh} in terms of real coefficients. 
We introduce the real spherical harmonics:
\begin{flalign}\label{eq:real_sh}
    {\cal Y}_{lm}= \begin{dcases}
    \frac{i}{\sqrt{2}} \left( Y_l^m  - (-1)^m Y_l^{-m} \right) & m<0\\
    Y_l^0 & m=0\\
    \frac{1}{\sqrt{2}} \left( Y_l^{-m}  + (-1)^m Y_l^m \right) & m>0\\
\end{dcases}
\end{flalign}
and rewrite the $l=1$ term of eq.~\eqref{eq:df_sh}, $\delta f_{1m}$, as:
\begin{equation}\label{eq:df_l1}
    \delta f (\bm{p})\Big|_{l=1} = {\cal Y}_{11} \delta f_x + {\cal Y}_{1-1} \delta f_y + {\cal Y}_{10} \delta f_z
\end{equation}
where the coefficients are given by
\begin{equation}\label{eq:df_phys}
    \delta  f_x=  \frac{\delta f_{1-1} - \delta f_{11}}{\sqrt{2}}, \quad \delta  f_y= \frac{ \delta f_{1-1} + \delta f_{11}}{\sqrt{2} i}, \quad \delta  f_z=  \delta f_{10}.
\end{equation}
Since the electron distribution function is real, so are the coefficients defined above. 
Substituting eq.~\eqref{eq:real_sh} and eq.~\eqref{eq:df_phys} in the collision integral eq.~\eqref{eq:I_sh}, we get
\begin{flalign}
     & I[f] =  \begin{pmatrix}
    {\cal Y}_{11} & {\cal Y}_{1-1} & {\cal Y}_{10} \\
    \end{pmatrix} \begin{pmatrix}
    \nu_{xx} & \nu_{xy} & \nu_{xz} \\
    \nu_{yx} & \nu_{yy} & \nu_{yz} \\
    \nu_{zx} & \nu_{zy} & \nu_{zz} \\
    \end{pmatrix}  \begin{pmatrix}
    \delta f_x \\
    \delta f_y \\
    \delta f_z \\
    \end{pmatrix} 
\end{flalign}
with the physical collision frequencies given by:
\begin{widetext}
\begin{equation}
\hat{\nu}=
\begin{pmatrix}
    \frac{1}{2} \left( \nu_{11}^{11} + \nu_{1-1}^{1-1} - \nu_{11}^{1-1} - \nu_{1-1}^{11} \right) 
    & \frac{i}{2} \left( \nu_{11}^{11} - \nu_{1-1}^{1-1} + \nu_{11}^{1-1} - \nu_{1-1}^{11} \right)
    & \frac{1}{\sqrt{2}} \left( \nu_{1-1}^{10} - \nu_{11}^{10}  \right)
\\
    \frac{i}{2} \left(- \nu_{11}^{11} + \nu_{1-1}^{1-1} + \nu_{11}^{1-1} + \nu_{1-1}^{11} \right)
    & \frac{i}{2} \left( \nu_{11}^{11} + \nu_{1-1}^{1-1} + \nu_{11}^{1-1} + \nu_{1-1}^{11} \right)
    & \frac{i}{\sqrt{2}} \left( \nu_{1-1}^{10} + \nu_{11}^{10}  \right) 
    \\
    \frac{1}{\sqrt{2}} \left( \nu_{10}^{1-1} - \nu_{10}^{11}  \right)
    & \frac{-i}{\sqrt{2}} \left( \nu_{10}^{11} + \nu_{10}^{1-1}  \right)
    &\nu_{10}^{10} \\
    \end{pmatrix}
\end{equation}
\end{widetext}
The constraint of elasticity eq.~\eqref{eq:W_el} implies that the collision frequency matrix is symmetric, $\nu_{ij} = \nu_{ji}$. Moreover, we can see from the pasta symmetries in eq.~\eqref{eq:W_sym} that the off-diagonal terms vanish and that the $xx$ and $yy$ terms are equal. 
This is valid for slabs because $L_{1x} = L_{1y}=L_1$. Thus,
\begin{equation}
\hat{\nu}=
\begin{pmatrix}
    \nu_{xx} & \nu_{xy} & \nu_{xz} \\
    \nu_{yx} & \nu_{yy} & \nu_{yz} \\
    \nu_{zx} & \nu_{zy} & \nu_{zz} \\
    \end{pmatrix} 
=
    \begin{pmatrix}
    \nu_{p} & 0 & 0  \\
    0 & \nu_{p} & 0 \\
    0 & 0  & \nu_{a} \\
    \end{pmatrix}
\end{equation}
where $\nu_p= \nu_{11}^{11}$ and $\nu_a= \nu_{10}^{10}$.   
Writing, without any loss of generality, 
$\delta f_{1m} = \sqrt{4\pi/3} \Phi_{1m}(\upepsilon_p) |\bm{v}|$, the collision integral expansion, eq.~\eqref{eq:I_sh}, can be {simply} rewritten as 
\begin{equation}\label{eq:I_yak}
    I[f] = - v_z  \Phi_z \nu_a - \bm{v}_p \cdot \bm{\Phi}_p \nu_p,
\end{equation}
where $\bm{\Phi}$ is a vector that can depend on $\upepsilon_p$, and the collision frequencies $\nu_a$ and $\nu_p$ are defined parallel and perpendicular to the pasta symmetry axis. This result exactly coincides with the generalization of the relaxation time approximation proposed by \citet{Yakovlev2016} on symmetry arguments to include the effect of the anisotropic medium. 

The axial and perpendicular collision frequencies can be calculated from eq.~\eqref{eq:nu_final}:
\begin{flalign}
    \nu_a (\upepsilon_p) &= \frac{p^2}{4 \pi^2 v} \bigg[ \W_{00, 00} -   \W_{10, 10} \nonumber\\
    & \hspace{2.5cm} +\frac{1}{\sqrt{5}} \left( W_{20, 00} + \W_{00, 20}\right) \bigg] \\
    \nu_p (\upepsilon_p) &= \frac{p^2}{4 \pi^2 v} \bigg[ \W_{00, 00} - \frac{1}{2\sqrt{5}} \left(\W_{20, 00} + \W_{00, 20} \right) \nonumber  \\
    & \hspace{2.5cm} + \frac{1}{2} \left( \W_{11, 1-1} +\W_{1-1, 11} \right) \bigg].
\end{flalign}

To rewrite $\nu_a$ and $\nu_p$ in terms of the transition matrix { $W_{pp'}$}, we invert eq.~\eqref{eq:Wexpansion} using the orthogonality of spherical harmonics:
\begin{equation}\label{eq:W_label}
    \W_{lm\,l'm'} = \int d\Omega_p d \Omega_{p'} W_{pp'} {{Y}_l^m} ^\ast (\Omega_p)  {{Y}_{l'}^{m'}}^\ast (\Omega_{p'}).
\end{equation}
This leads to the final expression of the collision rates, for an arbitrary interaction preserving axial symmetry, and assuming a dipole-like deviation from the equilibrium of the electron distribution function:
\begin{widetext}
\begin{flalign}
\nu_a\left(\upepsilon_p\right) &= \frac{3 }{ 32 \pi^3 v } \int d \Omega_p d \Omega_{p'}  W_{pp'} \, q^2\, \cos^2 \theta_q \label{eq:nua_general}\\
\nu_p\left(\upepsilon_p\right) &= \frac{3 }{ 32 \pi^3 v } \int d \Omega_p d \Omega_{p'} \, W_{pp'} \, q^2 \frac{1}{2}\sin^2 \theta_q. \label{eq:nup_general}
\end{flalign}
\end{widetext}
To get the generalization of  eqs~\eqref{eq:sigmakappa_iso} and \eqref{eq:nu_iso} to the physical problem of electron-pasta scattering, we now turn to evaluate the transition matrix $W_{pp'}$.

\section{Conductivity tensor for rods and slabs} 
\label{sec:nuap}

\subsection{Elastic scattering matrix in the incoherent scattering limit}
\label{sec:incoherent}

\begin{figure*}
    \includegraphics[scale=0.72]{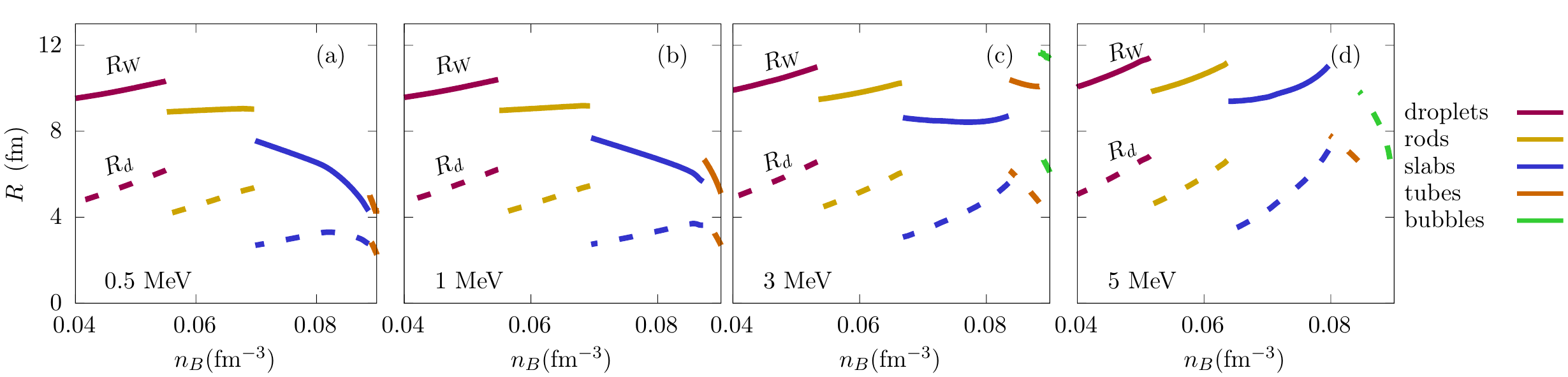}
    \caption{Linear and Wigner-Seitz radius of the pasta as a function of density in dashed and full curves for $T=0.5, 1, 3$ and 5 MeV (panels a, b c and d, respectively). The different pasta geometries are indicated with different colours. The results for $T=0$ and $T=0.5$ MeV are indistinguishable. }
    \label{fig:radius}
\end{figure*}

In the case of isotropic scattering, the transition matrix~\eqref{eq:Wexpansion}  depends solely on the absolute value of the transferred momentum  $\bm{q} = \bm{p} - \bm{p}\,'$. This is equivalent to a dependence on the angle between the incoming and outgoing electron momentum because one can write  $q^2 = 2 p^2 (1- \hat{p}\cdot \hat{p}\,')$.  
In the case of anisotropic scattering, the transition matrix can depend separately on the angles of both incoming and outgoing momenta. {However, because of the axial symmetry of the pasta structures}, it {can only} depend on the projections of the transferred momentum in the axis perpendicular and parallel to the symmetry axis of the pasta. We can assume without loss of generality that the pasta symmetry axis coincides with the $z$ axis, such that the transition matrix becomes a function of the vector $\bm{q}$ itself, {see Fig.\ref{fig:pasta}}.

The transition matrix  $W_{pp'}$ for the elastic scattering in the Born approximation is given by (see, e.g., equation 81.5 of~\citealt{berestetskii1971relativistic})
\begin{flalign}
    W_{pp'}(\bm{q}, \upepsilon_p) = &\frac{1}{2} \sum_{s, s'} \left|\frac{e}{2 \upepsilon_p} \bar u_{p', s'} \gamma^0 u_{p, s} \int d^3 \bm{x}   A_0(\bm{x}) e^{-i \bm{q} \cdot \bm{x} } \right|^2\nonumber \\
    &= e^2 \left( 1- \frac{q^2}{4 \upepsilon_p^2}\right)\left|  U(\bm{q}) \right|^2 S(\bm{q}) \label{eq:W_scattering}
\end{flalign}
where in the first line $u_{p, s}$ is the electron spinor, $A_0(\bm{x})$ is the electric potential generated by nuclei and $\gamma^0$ is the Dirac matrix. In the second line, the Fourier transform of the potential $U(\bm{q})$ is introduced, and the  static structure factor is defined as~\citep{flowers1976}:
\begin{flalign}\label{eq:Sq}
    S(\bm{q}) & =  
    \langle n_p(\bm{q})n_p(-\bm{q}) \rangle_T  \nonumber \\
    &= \frac{1}{V} \int d^3\bm{ r}d^3\bm{ r'}  e^{i\bm{q}\cdot\left( \bm{r} - \bm{r'} \right)} \langle  n_p(\bm{r}) n_p(\bm{r'})\rangle_T,
\end{flalign}
where $n_p(\bm{q})$ is the charge density of the scatterer in momentum space, $\langle ...\rangle_T$ is the thermal average that accounts for correlations between protons, and the integral covers the entire  thermodynamic system of volume $V$.
The structure factor carries the whole information regarding the anisotropy of the system both through the anisotropic shape of the pasta and through the lattice arrangement. 
In principle, $S(\bm{q})$ also carries information about thermal excitations. The contribution of single-nucleon thermal excitations to $S(\bm{q})$ has been calculated by~\citet{PhysRevC.101.055804}, within the framework of density functional theory. Since this is not a main source of anisotropy, we do not consider it here. On the other hand, larger-scale collective thermal vibrations of pasta structures and deviations from lattice periodicity are likely important to the anisotropic behaviour of transport coefficients. To the best of our knowledge, these have not been calculated yet and will be addressed in future work.
\begin{figure*} 
\centering
\includegraphics[scale=0.75]{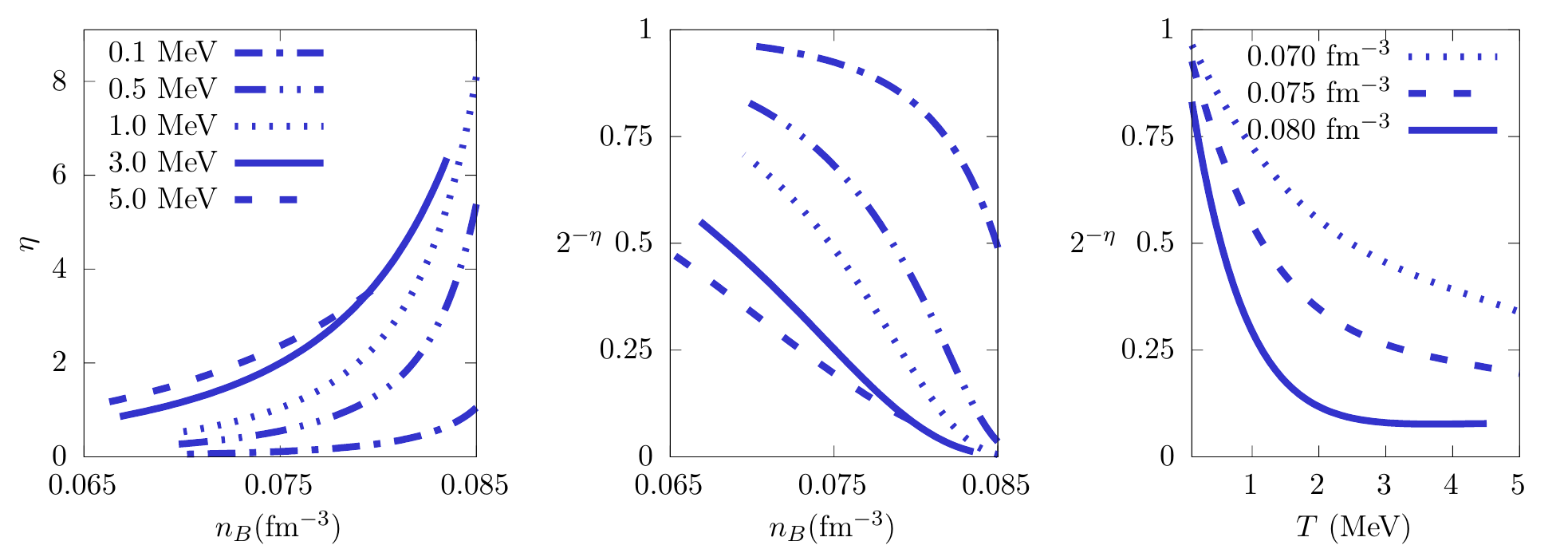}
\caption{ Left panel: correlation function decay parameter $\eta$ for slabs. Middle (right) panel: estimation of the correlation between neighbouring slab structures   as a function of density (temperature), see text for details. Curves  for $T=$0.1, 0.5, 1, 3, and 5 MeV are shown on the left and center, and for $n_B=$ 0.07, 0.075 and 0.08 fm$^{-3}$ on the right.}
    \label{fig:eta}
\end{figure*}

Still, variational theory in Wigner-Seitz (WS) cells of different geometries with energy densities obtained  from the RMF approach, is routinely used by nuclear physicists to obtain a microscopically funded estimation of the optimal average charge distribution $\langle  n_p(\bm{r}) \rangle_T$, see e.g. \citet{Avancini:2008kg,PhysRevC.78.015802,PRC85} and \citet{haensel2007neutron,chamel_2008LRR} for reviews.
In the simplest liquid-drop modelling of Fig.~\ref{fig:pasta}, the pasta structures are characterized by constant density profiles. 
We can write for rods ($d=2$) and slabs ($d=1$), respectively:
\begin{equation}
\label{pippo}
\begin{split}
& \langle n_{p}(\bm{r})\rangle_T^{d=2}=n_{p}\sum_m \Theta(R_2 - r-m R_{W2} ) 
 \\
 & \langle n_{p}(\bm{r})\rangle_T^{d=1}=n_{p}\sum_m \Theta(R_1 - z- m R_{W1})
\end{split}
\end{equation}
where the average linear size of the cluster $R_d$, its internal proton (neutron) density $n_p$ ($n_n$) and average WS cell radius $R_{Wd}$ are variationally obtained for any given temperature $T$ and baryon density $n_B$, as well as the (uniform) electron density, and the density of the dripped neutrons\footnote{
    This last quantity does not play any role in the calculations concerned by the present paper.
}. 
The sum in \eqref{pippo} runs over the parallel structures in the lattice. For this application, we use the relativistic mean-field approach of~\citet{PRC85}.
The mean field Lagrangian is given by the non-linear Walecka model, parametrized by the IUFSU force~\citep{IUFSU} with a surface tension fitted to reproduce a Thomas-Fermi simulation, see ~\citet{PRC85} for details. 
In Fig.~\ref{fig:radius} we show the linear and WS cell radii as a function of the density, for some representative temperatures that will be considered below.
 We can see that non-spherical geometries are expected in the innermost part of the inner crust, and are found to persist even at high temperatures well above crustal melting, that occurs around $T_m\approx 1$ MeV in this density region~\citep{Carreau2020}.  
The different colours correspond to the geometries that are associated, for a given baryon density, with the minimal free energy density.
It is known that the pasta properties are model dependent~\citep{DinhThi:2021bxj}, mainly the densities at which the different geometries appear, but the qualitative behaviour shown in Fig.\ref{fig:radius} is obtained in all realistic nuclear models found in the literature~\citep{DinhThi:2021bxj,PhysRevD.106.023031}.

In the case of a perfect lattice, electron band structures suppress the scattering rates and charge transport occurs only through electron-phonon interactions. However, thermal fluctuations disturb the lattice periodicity and destroy the long-range order. In particular, in the disordered limit expected to dominate with increasing temperature, the correlation function drops to zero on a length scale of the linear size of the cluster, and the different pasta structures are fully uncorrelated and act as incoherent impurity scatterers.   
These fluctuations have been calculated for the pasta by~\citet{PETHICK19987} within the classical approach of the Landau-de Gennes model of liquid crystals~\citep{DeGennes,Chandrasekhar}.  For slabs, the thermal displacement presents a logarithmic divergence with the linear dimension of the sample, reflecting the well-known Landau-Peierls instability~\citep{Landau}. Concerning the rod phase, the thermal displacement is finite and the long-range order in the transverse plane is in principle preserved. 
A critical temperature for the long (or quasi-long) range order was estimated by \citet{WATANABE2000455} as the temperature at which the thermal displacement becomes comparable to the cell radius. Such a temperature was shown to strongly decrease with increasing baryonic density and, for fiducial values of the elastic constants, to be of the order of a few MeV both for slabs and for rods \citep{WATANABE2000455}. Above such temperatures, it is reasonable to expect that the conventional pasta phase is fully destroyed by thermal fluctuations, even if complex deformed disordered cluster structures may still be present, as suggested by molecular dynamics simulations ~\citep{Horowitz_prl2015, PhysRevC.90.055805,newton_glassy_pasta,Nandi_2018}. Below the critical temperature, the consequence of the reduced dimensionality of the pasta phase is that the long-range order (or quasi-long in the case of slabs) is only preserved in the directions corresponding to the lattice periodicity (that is along $\bm{u}_z$ for the slab phase, and  $\bm{u}_\perp$ for the rod one), potentially leading to strong anisotropies in the collision frequencies.

\begin{figure*}
    \centering
    \includegraphics[scale=.8]{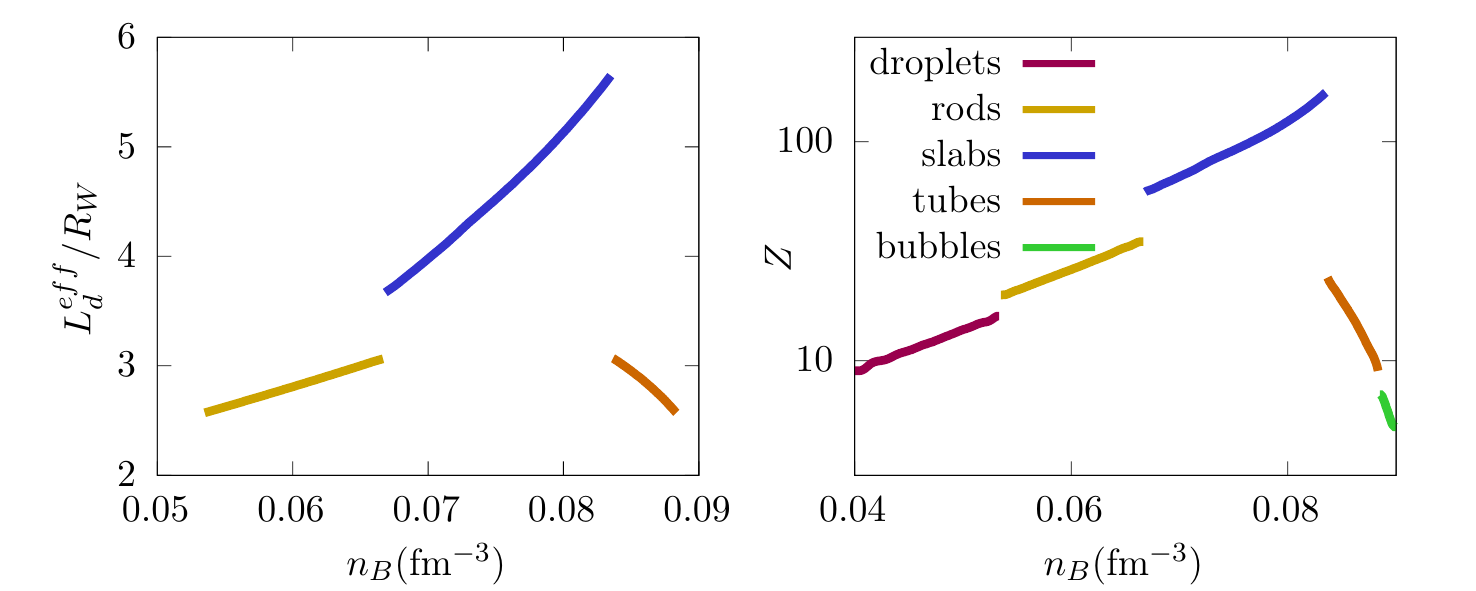}
    \caption{Left: Effective length of the pasta transverse to the symmetry axis, normalized to the Wigner-Seitz radius (see text). Right: Proton number of the pasta as a function of density. The different pasta geometries are indicated with different colours. The temperature is fixed to  $T=3$ MeV. }
    \label{fig:length}
\end{figure*}

Interestingly, limiting behaviours can be obtained for the density correlation of slabs  \citep{Poniewierski98,DeGennes} $\langle  n_p(\bm{r}) n_p(\bm{r'})\rangle_T\equiv\langle \delta n^2(\bm{r}-\bm{r}')\rangle_{T}$ showing the power law behaviour characteristic of the quasi-long range order of the smectic (slab) phase:
\begin{eqnarray}\label{eq:deltan_slab}
    \langle \delta n^2(\bm{r})\rangle_{T} &\propto& |z| ^{-\eta}  
      \;\; |z|\to\infty \\
     &\propto&   r_\perp ^{-2\eta}  \;\; r_\perp \to\infty , \label{eq:deltan2_slab}
\end{eqnarray}
where 
\begin{equation}
    \eta=\frac{q_0^2T}{8\pi\lambda C_0},
\end{equation}
with $q_0=  \pi/R_{W1}$, $\lambda^2= R_{W1}^2 (1 + 2f - 2 f^2)/45$ where $f=R_1/R_{W1}$ is the average volume fraction of the cluster,and $C_0= 6 E_C$, where $E_C$ is the average Coulomb energy density in the cell~\cite{PETHICK19987}. More recently, the calculations of elastic constants were improved by~\citet{Pethick:2020aey}, but for our estimates we stick to the simpler prescription of~\citet{PETHICK19987}. 
The numerical values of the $\eta$ parameter as a function of the density, as numerically obtained from the average pasta configuration predicted by the RMF model, are displayed in the left part 
of Fig.~\ref{fig:eta} for different temperatures. As expected, the correlation decreases with temperature and density.

In the absence of a complete calculation of the correlation function, we limit ourselves in this paper to temperatures high enough for the hypothesis of uncorrelated scatterers to be realistic. To this aim, we plot in the center and right parts of Fig.~\ref{fig:eta} the quantity $2^{-\eta}$ as an estimation of the ratio between the correlation function at $z=2R_{W1}$, corresponding to a distance containing two different slabs, and the same quantity at  $z=R_{W1}$, such that a single slab is accounted for, 
\begin{equation}\label{eq:eta1}
    2^{-\eta}\approx\frac{\delta n^2 (2R_{W1})}{\delta n^2 (R_{W1})}.
\end{equation}
Even if these distances might be small to justify the use of the asymptotic behaviour given by eq.~\eqref{eq:deltan_slab}, the quantity $2^{-\eta}$ can be taken as an estimation of the correlation reduction due to thermal effects.
From Fig.\ref{fig:eta} we can see that only at very high temperatures above 1 MeV the hypothesis of incoherent scattering appears justified. For the following numerical applications, we will focus on $T=3$ MeV as a representative temperature value.

Since the correlation asymptotically follows the same power law in the transverse as in the longitudinal plane see eq.~\eqref{eq:deltan2_slab}, we define the effective length of the slab $L_1$ from the same order-of-magnitude consideration:
\begin{equation}\label{eq:eta2}
    \frac{\delta n^2 (L_1^{eff})}{\delta n^2 (L_{W1})} \approx  2^{-\eta}
\end{equation}
where $L_{W1}$ is defined by normalizing the slab WS volume to the droplet volume at identical thermodynamic conditions. Comparing eqs \eqref{eq:eta1} and \eqref{eq:eta2}, we consider that the effective length of the slabs is 
\begin{equation}
    L_1^{eff} = \sqrt{2} L_{1W}.
\end{equation}
For the rods, in the absence of an exact calculation of $S(\bm{q})$, we assume the length of interest to be equal to the slabs one if they were dominant at the same density, $L_2^{eff}=L_1^{eff}$.
The resulting numerical values of the pasta length and proton number at $T=3$ MeV are shown in Fig.~\ref{fig:length}.

Within the hypothesis of incoherent scatterings, the  structure factor integral is limited to a single cell. 
Charge fluctuations within the cell being negligible we can write
$\langle  n_p(\bm{r}) n_p(\bm{r'})\rangle_T=\langle  n_p(\bm{r})\rangle_T \langle n_p(\bm{r'})\rangle_T$, leading to:
\begin{equation}\label{eq:Sq_final}
    S(\bm{q})\approx  Z^2 n_i |F(q)|^2,
\end{equation}
with the form factor $F(\bm q)$ defined as the Fourier transform of the charge density normalized by the number of protons composing the cluster ($Z$):
\begin{equation}\label{eq:ff_def}
    F(\bm{q}) = \frac{1}{Z} \int_{\rm WS} d^3 \bm{r} e^{i \bm{q} \cdot \bm{r} } n_{p}(\bm{r}), 
\end{equation}
and the number density of  targets within the medium as $n_i=1/V$.  

Analytic expressions can be found for the form factor by direct integration of eq.~\eqref{eq:ff_def} for spherical, cylindrical, and planar geometries  (labelled $d=3, 2, 1$, respectively):
\begin{widetext}
\begin{equation}\label{eq:form_factors}
    F_d(\bm{q}) = \left.
\begin{dcases}
 \frac{3}{(q R_3)^3} \left[ \sin(qR_3) -qR_3 \cos(qR_3) \right], & \text{d=3 }\\
\frac{2 }{q_z L_2} \sin\left( \frac{q_z L_2 }{2}\right) \frac{2 }{q_\perp R_2 } J_1\left( q_\perp R_2\right), & \text{d=2}\\
\frac{2}{L_1 q_x} \sin\left( \frac{L_1 q_x}{2}\right) \frac{2}{L_1 q_y} \sin\left( \frac{L_1 q_y}{2}\right)\frac{1}{R_1 q_z} \sin\left( {R_1 q_z}\right), & \text{d=1}
\end{dcases}
\right.
\end{equation}
\end{widetext}
Here, $q_x= q \sin \theta_q \cos\phi_q$, $q_y=q\sin\theta_q\sin\phi_q$, $q_z=q\cos\theta_q$, $q_\perp = \sqrt{q_x^2 + q_y^2}$, and $J_1$ is the cylindrical Bessel function,
\begin{equation}
J_1(x)=\frac{1}{i\pi} \int_0^\pi d\phi e^{ix cos(\phi)} cos(\phi).
\end{equation}
Finally, using the Fourier transform of the electric potential
\begin{equation}
    U(\bm{q}) = \frac{4 \pi e}{q^2 \varepsilon(\bm{q})} ,
\end{equation}
the transition matrix element is written  as:
\begin{equation}
    W_{pp'}(\bm{q}, \upepsilon_p) = n_i e^2 \left( 1- \frac{q^2}{4 \upepsilon_p^2}\right)\left| \frac{4 \pi e Z F_d(\bm{q})}{q^2 \varepsilon(q)} \right|^2, \label{eq:W_pasta}
\end{equation}
where the dielectric function $\varepsilon(q)$ is introduced, regularizing the divergence at $q=0$, to account for electron screening. A complete calculation within relativistic quantum mechanics, in the random phase approximation, gives~\citep{jancovici1962relativistic, haensel2007neutron}:
\begin{flalign}
    \epsilon(q) &= 1 + \frac{k_{TF}^2}{q^2 } \Bigg\{ \frac{2}{3} - \frac{2}{3} \frac{y^2 x_r}{\gamma_r} \ln(x_r + \gamma_r) \nonumber \\
    & + \frac{x_r ^2 +1 - 3 x_r^2 y^2 }{6 y x_r^2} \ln \left| \frac{1+y}{1-y} \right| \nonumber \\
    & + \frac{2y^2 x_r^2 -1}{6 y x_r^2} \frac{\sqrt{1+x_r^2 y^2}}{\gamma_r} \ln \left| \frac{y \gamma_r + \sqrt{1+x_r^2y^2}}{y \gamma_r - \sqrt{1+x_r^2y^2}} \right| \Bigg\},
\end{flalign}
where $y = q/2 p_F$, $x_r= p_F/m_e$,  $\gamma_r= \sqrt{1+x_r^2}$, 
$k_{TF}$ is the Thomas-Fermi momentum, defined as:
\begin{equation}
    k_{TF} = \sqrt{4\pi e^2 \partial n_e / \partial \mu_e} 
    =2 \sqrt{\alpha_{em} \gamma_r/(\pi x_r)} k_F,
\end{equation}
and the second equality supposes a strongly degenerate electron gas ~\citep{haensel2007neutron}. Though in this work we assume that the screening is isotropic, it is important to observe that strong magnetic fields lead to anisotropic behaviour and can produce Friedel oscillations~\citep{PhysRev.186.434, PhysRevC.83.025803}. We leave the consideration of this source of further anisotropies to a future study. 

\begin{figure*}
    \centering
    \includegraphics[scale=0.8]{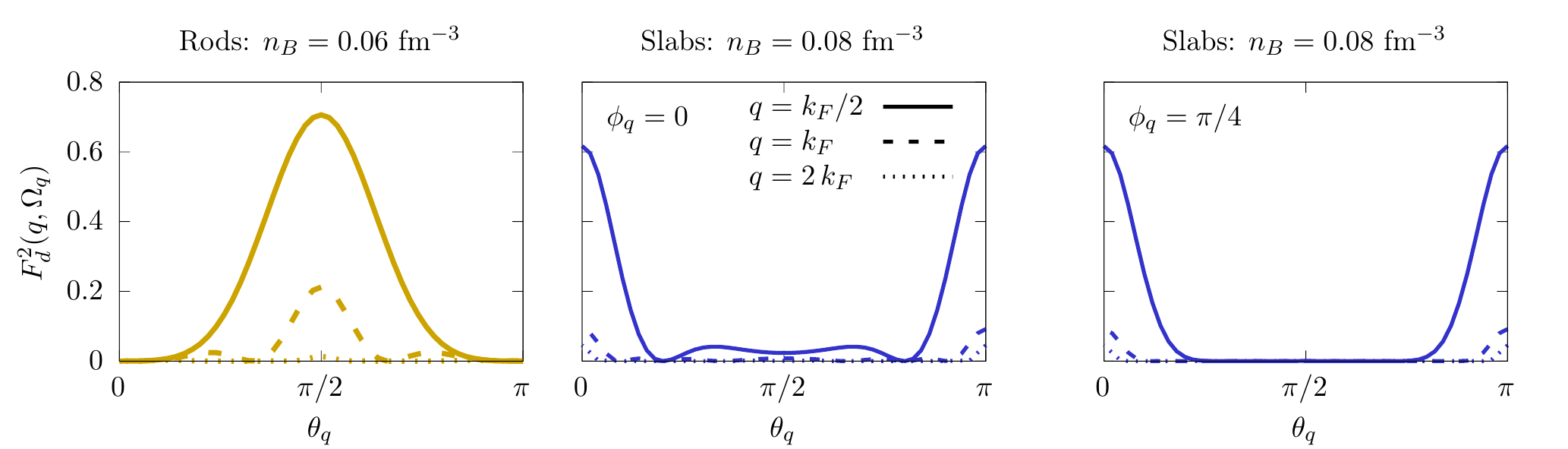}
    \caption{{Form factor squared of rods at the representative density  $n_B=0.06$ fm$^{-3}$ (left, yellow) and slabs at  $n_B=0.08$ fm$^{-3}$  (middle and right, blue)   as a function of the azimuthal angle, for different transferred momenta $q=k_F/2, k_F$ and $2 k_F$ shown as continuous, dashed and dotted curves, respectively.} For slabs we fix $\phi_q=0$ (middle) and $\pi/4$ (right). The effective lengths $L$ are taken from Figure \ref{fig:length}. The  temperature is fixed to  $T=3$ MeV.}
    \label{fig:Ftheta_}
\end{figure*}

\subsection{Collision frequencies and conductivities} \label{sec:tensor}

The expression of the transition matrix in eq.~\eqref{eq:W_pasta} allows us to get the final result for the collision rates. We substitute this expression into eqs~\eqref{eq:nua_general} and~\eqref{eq:nup_general} and make the change of variables
\begin{equation}\label{eq:change_var}
    \int d \Omega_p d \Omega_p'= \frac{2 \pi}{p^2} \int  \frac{d^3\bm{q}}{q} \, ,
\end{equation}
to get the collision frequencies
\begin{widetext}
\begin{flalign}
\nu_a\left(\upepsilon_p\right) &= \frac{12 \pi n_i e^4 Z^2}{v p^2} \int_0^{2 p} dq \frac{1}{q \varepsilon^2(q)}\left( 1- \frac{q^2}{4 \upepsilon_p^2}\right) \int \frac{d \Omega_q}{4 \pi}  \left|F_d(\bm{q})\right|^2  \cos^2 \theta_q \label{eq:nua}\\
\nu_p\left(\upepsilon_p\right) &= \frac{12 \pi n_i e^4 Z^2}{v p^2} \int_0^{2 p} d q \frac{1}{q \varepsilon^2(q)}\left( 1- \frac{q^2}{4 \upepsilon_p^2}\right) \int \frac{d \Omega_q}{4 \pi}\left|F_d(\bm{q})\right|^2 \frac{1}{2}  \sin^2 \theta_q \, . 
\label{eq:nup}
\end{flalign}
\end{widetext}
Eqs~\eqref{eq:nua} and \eqref{eq:nup} are the main results of the paper. 
Under the assumption of incoherent scatterings among the different pasta structures, and
using the analytical expressions eq.~\eqref{eq:form_factors} for the form factors, they allow calculating the thermal and electric conductivity of the pasta phase (see eq.~\eqref{eq:sigma} below) at any arbitrary temperature, density, proton fraction and magnetic field value from a given nuclear physics model providing the composition of the matter, namely the values of $L$, $R$ and $Z$ for the dominant pasta geometry. 
Some representative results will be given in the next section.

To compute the transport coefficients, the collision frequencies eqs~\eqref{eq:nua} and \eqref{eq:nup} must be calculated at the Fermi energy $\upepsilon_p=\upepsilon_F$, since in the strongly degenerate electron gas transport occurs close to the Fermi surface. 
They can be written compactly as 
\begin{equation}\label{eq:nuap}
    \nu_K = \frac{12 \pi n_i e^4 Z^2}{v_F p_F^2} \Lambda_{K} ,
\end{equation}
where $p_F$ ($v_F$) is the Fermi momentum (velocity). We have also defined the axial ($K=a$) and perpendicular ($K=p$) Coulomb logarithms as
\begin{equation}\label{eq:LambdaK}
    \Lambda_K= \int_0^{2 p_F} \frac{d q}{q} \frac{1}{ \varepsilon^2(q)}\left( 1- \frac{q^2}{4 \upepsilon_F^2}\right) \langle F^2 \rangle_K \, ,
\end{equation}
and the averages $\langle F^2 \rangle_K$ as
\begin{equation}\label{eq:Fa}
\langle F^2 \rangle_a =
\int \frac{d \Omega_q}{4 \pi}  \left|F_d(\bm{q})\right|^2  \cos^2 \theta_q \, ,
\end{equation}
and 
\begin{equation}\label{eq:Fp}
\langle F^2 \rangle_p = 
\int \frac{d \Omega_q}{4 \pi}  \left|F_d(\bm{q})\right|^2 \frac{1}{2} \sin^2 \theta_q \, .
\end{equation}
The calculation of the conductivities with the anisotropic collision frequencies has been worked out in~\citet{Yakovlev2016}, so here we only report the main equations. The magnetic field is assumed, without loss of generality, to lie in the $xz$--plane. 
By defining the unit vector $\bm{b} = \bm{B}/B = b_x \hat{x} + b_z \hat{z}$ and the cyclotron frequency for electrons  
$\omega= e B/\upepsilon_F$,  the electric conductivity tensor can be written as
\begin{equation} \label{eq:sigma}
    \hat{\sigma} = \frac{e^2 n_e}{ m_e^\ast \Delta} 
    \begin{pmatrix}
    \nu_a \nu_p + \omega^2 b_x^2 &  - \omega b_z \nu_a  & \omega^2 b_x b_z \\
    \omega b_z \nu_a             &  \nu_a \nu_p         & - \omega b_x \nu_p \\
    \omega^2 b_x b_z             &  \omega b_x \nu_p    & \nu_p^2 + \omega^2 b_z^2 
    \end{pmatrix} ,
\end{equation}
where $\Delta = \nu_p^2 \nu_a + \omega^2 b_x^2 \nu_p + \omega^2 b_z^2 \nu_a$. The thermal conductivity can be obtained by the Wiedemann-Franz law $\kappa_{ij} = \sigma_{ij} (\pi^2 T/3e^2)$ which is valid for strongly degenerate electrons~\citep{ziman2001electrons}. 
We do not consider neutron superfluidity, which results in corrections to the thermal conductivity, but not to the electrical~\citep{PhysRevLett.102.091101}. 
For $B=0$, the conductivity becomes
\begin{equation}
\label{eq:sigmaK}
    \hat{\sigma}_0 = \frac{e^2 n_e}{ m_e^\ast} 
    \begin{pmatrix}
    \nu_p^{-1}  &  0            & 0 \\
    0           &  \nu_p^{-1}   & 0 \\
    0           &  0            & \nu_a^{-1}
    \end{pmatrix} .
\end{equation}

\section{Numerical results} 
\label{sec:pasta}

\begin{figure*}
    \centering
    \includegraphics[scale=0.8]{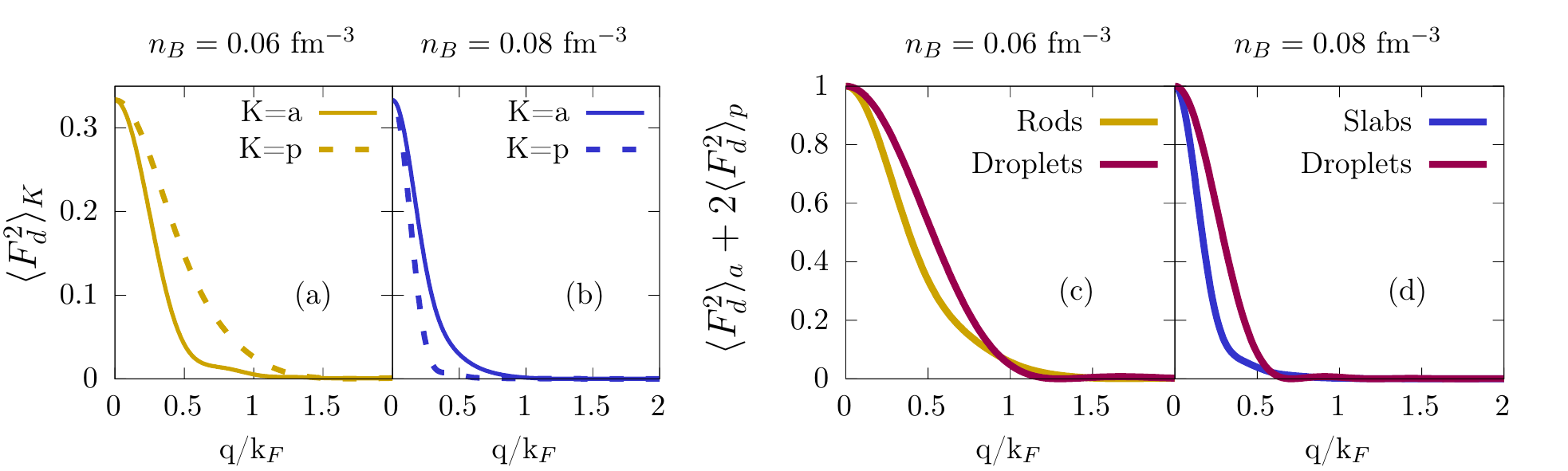}
    \caption{Left panels (a,b): axial (full) and perpendicular (dashed) square form factor of rods (yellow) and slabs (blue) at  $n_B=0.06$ (a) and $0.08$ (b) fm$^{-3}$. Right panels (c,d): average square form factors of rods (yellow, c panel) and slabs (blue, d panel) are  compared to the ones of droplets (magenta) at the same density. The representative temperature T=3 MeV is chosen.}
    \label{fig:F_}
\end{figure*}

The  source of anisotropy entering the collision frequencies eq. \eqref{sec:nuap} lies in the angular dependence of the form factors $F_d(\bm q)$. The latter is displayed for rods and slabs in Fig. \ref{fig:Ftheta_}, as a function of the azimuthal angle $\theta_q$ (see Fig.\ref{fig:pasta}) for different values of $q=k_F/2,k_F,2k_F$.  
In these figures, the temperature is fixed to $T=3\,$MeV and the two representative densities $n_B=$0.06, 0.08 fm$^{-3}$ are chosen, where rods (slabs) are expected to be dominant according to the results presented in Fig. \ref{fig:radius}.
We can see from Fig. \ref{fig:Ftheta_} that for the rods, the form factor is strongly  peaked at $\theta_q=\pi/2$, while for slabs it is peaked at $\theta_q = 0$ and $\pi$.  Such behaviour is expected from their geometries, as the form factors are peaked in the elongated direction. 

However, this dependence is smoothed out by the angular average implied by eqs~\eqref{eq:Fa} and \eqref{eq:Fp}. 
This is shown in Fig.~\ref{fig:F_}, that displays the averaged form factors $\langle F^2\rangle_{a,p}$ defined in  eqs~\eqref{eq:Fa},\eqref{eq:Fp}. 
The average value, given by $2\langle F^2 \rangle_p + \langle F^2 \rangle_a $, is also plotted, in the same thermodynamic conditions as in Fig. \ref{fig:Ftheta_}.
As we can expect from Fig.\ref{fig:pasta} and eq.~\eqref{eq:form_factors}, the form factor is maximum (minimum) in the symmetry axis direction in the case of slabs (rods). This difference is pronounced at momentum transfers smaller than $q=k_F$, as afterwards, both axial and perpendicular components tend to zero. When comparing the form factors to the one of the equivalent spherical droplet (right side of Fig.\ref{fig:F_}), we can see that the difference is essentially seen at low momentum as well, where the form factor is systematically smaller within a deformed shape  than for the equivalent spherical geometry.

\begin{figure}
    \centering
    \includegraphics[scale=.9]{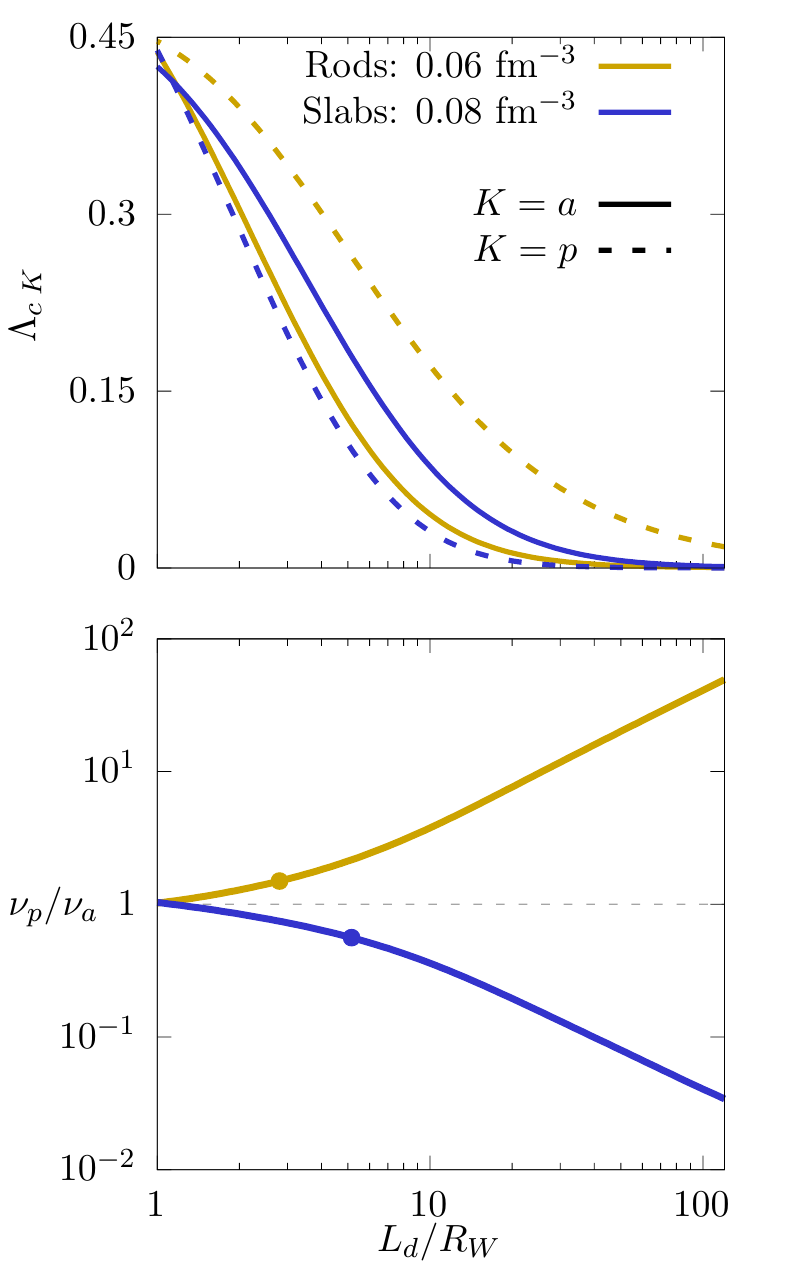}
    \caption{{The Coulomb logarithms (upper part) and the ratio of perpendicular to axial collision frequency (lower part) are shown as a function of the pasta length normalized to the Wigner-Seitz radius, $L_d/R_W$. Quantities for rods (slabs) are  calculated at $n_B=0.06$ fm$^{-3}$ ($n_B=0.08$ fm$^{-3}$) and plotted in yellow (blue). In the upper panel, the perpendicular (axial) component are displayed as dashed (continuous) lines.    } {The points in the lower plot indicate the estimated effective length  $L_1^{eff}= \sqrt{2}\, L_{1W}$ (see text)}.The representative temperature $T=3$ is chosen.} \label{fig:coul_log}
\end{figure}

In the previous figures, we have estimated the effective pasta length $L_d$ based on the asymptotic behaviour of the thermal density correlation function of the smectic phase in the perpendicular plane (see Figure \ref{fig:eta} and associated discussion). Though qualitatively the physical origin of the electron-pasta scattering is certainly the breaking of the long-range order due to thermal fluctuations, our estimations are very rough and would deserve to be confronted with microscopic molecular dynamics simulations \citep{Horowitz_prl2015,2021PhRvC.103e5810C,newton_glassy_pasta,PhysRevC.104.L022801,Nandi_2018}. To evaluate the effect of the uncertainty on the estimation of $L_d$, in Fig.~\ref{fig:coul_log} we show the axial and perpendicular Coulomb logarithms eq.\eqref{eq:LambdaK}  and the ratio of perpendicular to axial collision frequency eq.\eqref{eq:nuap} as a function of the ratio of the pasta length $L_d$ to the WS-radius $R_{Wd}$ for rods and slabs. 

\begin{figure*} 
\centering
	\includegraphics[scale=0.8]{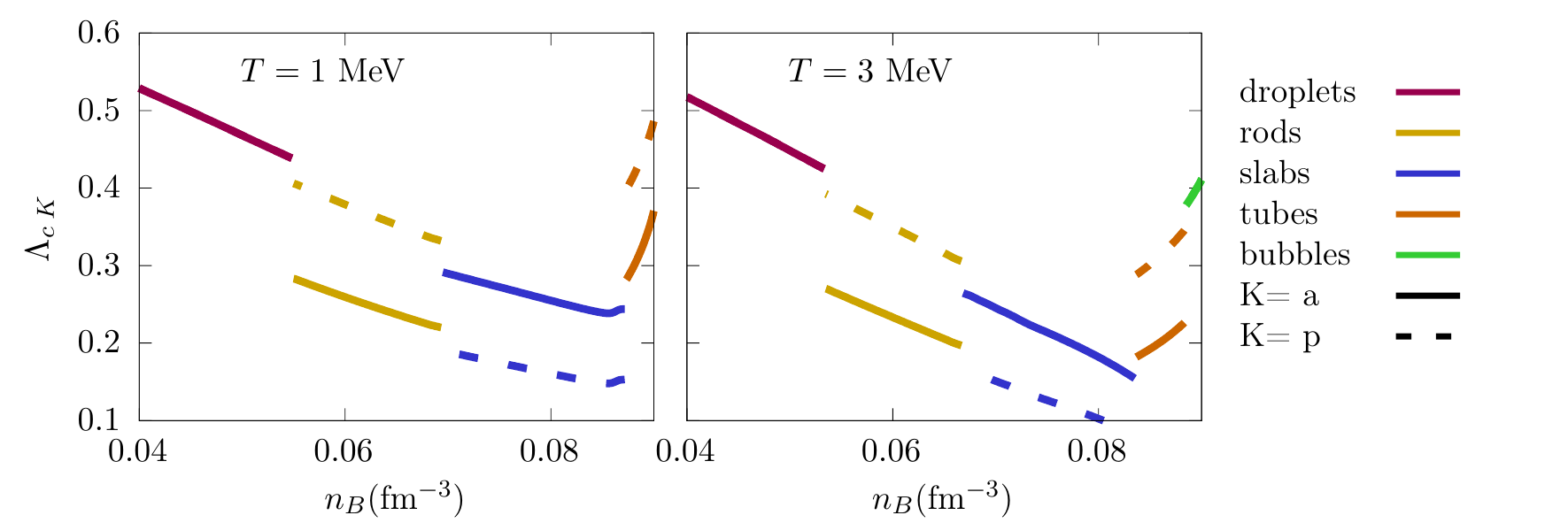}
\caption{Axial (continuous) and perpendicular (dashed) Coulomb logarithms as a function of baryon density, for $T=1$ MeV (left panel) and $T=3$ MeV (right panel). The different pasta geometries are indicated with different colours.  }
    \label{fig:coul_log_nb}
\end{figure*}

\begin{figure} 
\centering
	\includegraphics[scale=.9]{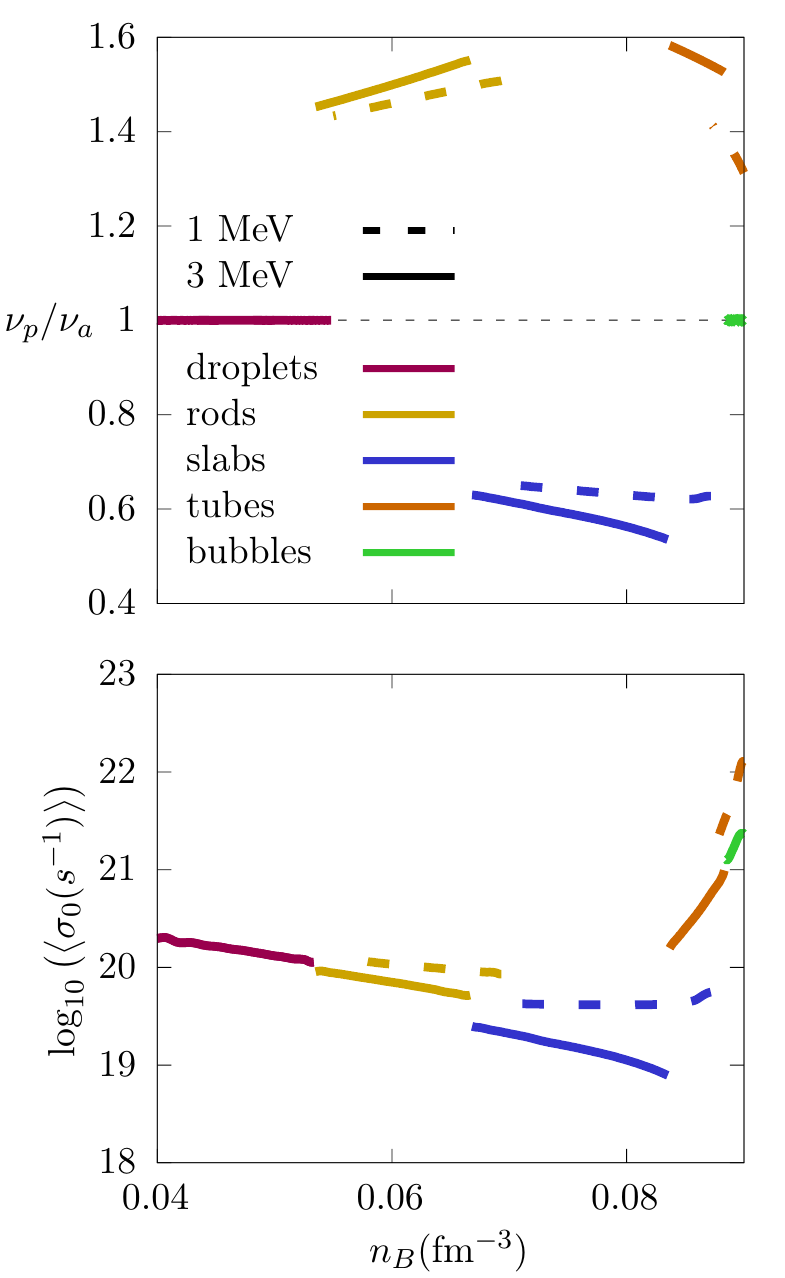}
\caption{Top: ratio of perpendicular to axial relaxation time. Bottom: Average electric conductivity. Curves are shown as a function of density for the representative temperatures of $T=$ 1 (dashed) and 3 (continuous) MeV. The different pasta geometries are indicated with different colours.}
    \label{fig:sigma_nb}
\end{figure}

For both geometries, the Coulomb logarithm decreases with  the increasing length of the pasta, and its value tends to
zero as $L_d$ becomes sufficiently large. 
This is consistent with the expectation that the lattice order should suppress the electron-ion scattering and increase the  conductivity of matter.

For rods (slabs), the perpendicular component is larger (smaller) than the axial one, and the difference between them increases with the growing length of the pasta, varying up to 100 (0.01) when $L_d \approx 150 R_W$.
We can see that a precise estimation of the length of the structures is important for the quantitative determination of the collision frequencies, as it affects in a considerable way the difference between the two scattering directions. In particular, the deviation from an isotropic scattering is small only for small values of $L_d/R_W$, corresponding to the high-temperature regime. At smaller temperatures, as correlations become more important, a larger transverse length will contribute to the scattering, so the difference in the anisotropic frequencies  will be more pronounced, likely reaching those expected in~\citet{Yakovlev2016}. 

{In Fig.~\ref{fig:coul_log_nb} the Coulomb logarithms  eqs\eqref{eq:LambdaK} are shown as a function of the density for T=1 and 3 MeV. In both cases,} 
we can see that the abrupt change of favoured geometry leads to slight discontinuities in the Coulomb logarithms, and both overall  decrease with density. This can be understood from the increase in length, shown in Fig.~\ref{fig:length}, and from Fig.~\ref{fig:coul_log}.

The ratio of perpendicular to axial collision frequency  eqs~\eqref{eq:nuap} and the
average conductivity are shown in Fig.~\ref{fig:sigma_nb} for the same temperatures. The slight increase in the ratio with density is due to the increasing length. It is important to note that, at the temperatures and lengths we are considering, the different collision frequencies differ by a factor smaller than two, so there is only a small deviation from isotropic scattering at high temperatures. 
 In the average conductivity on the other hand, there is a larger discontinuity when the abrupt change of geometry happens. This difference is mainly due to the associated discontinuity in proton number, which can be seen on the right side of Fig.~\ref{fig:length}. 
All in all, 
the anomalously high resistivity of the pasta layer reported in the literature ~\citep{PhysRevLett.121.132701} is nicely reproduced by our calculations, and it is seen to be essentially due to the high $Z$ value of the clusters close to the crust-core transition, more than to the specific geometry of the pasta phases.

\begin{figure*}
    \centering
    \includegraphics[scale=.9]{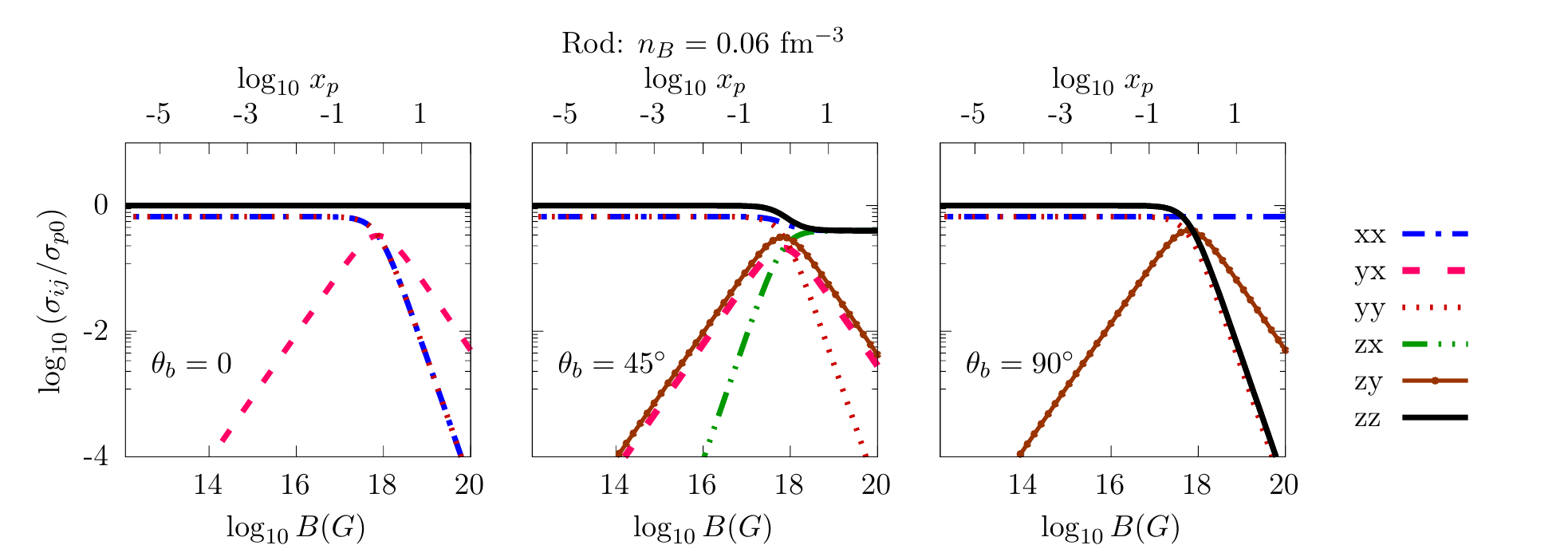}
      \caption{Components of the electric conductivity in units of the perpendicular conductivity at $B=0$, as a function of the magnetic field for rods at $n_B=0.06$ fm$^{-3}$.   The angle between the pasta symmetry axis and the magnetic field is fixed at 0 (left), 45º (center) and 90º (right). In the top axis, we show the variable $x_p = eB/(\upepsilon_F\nu_p)$. }
    \label{fig:sigma_x2}
\end{figure*}

 \begin{figure*}
    \centering
    \includegraphics[scale=.9]{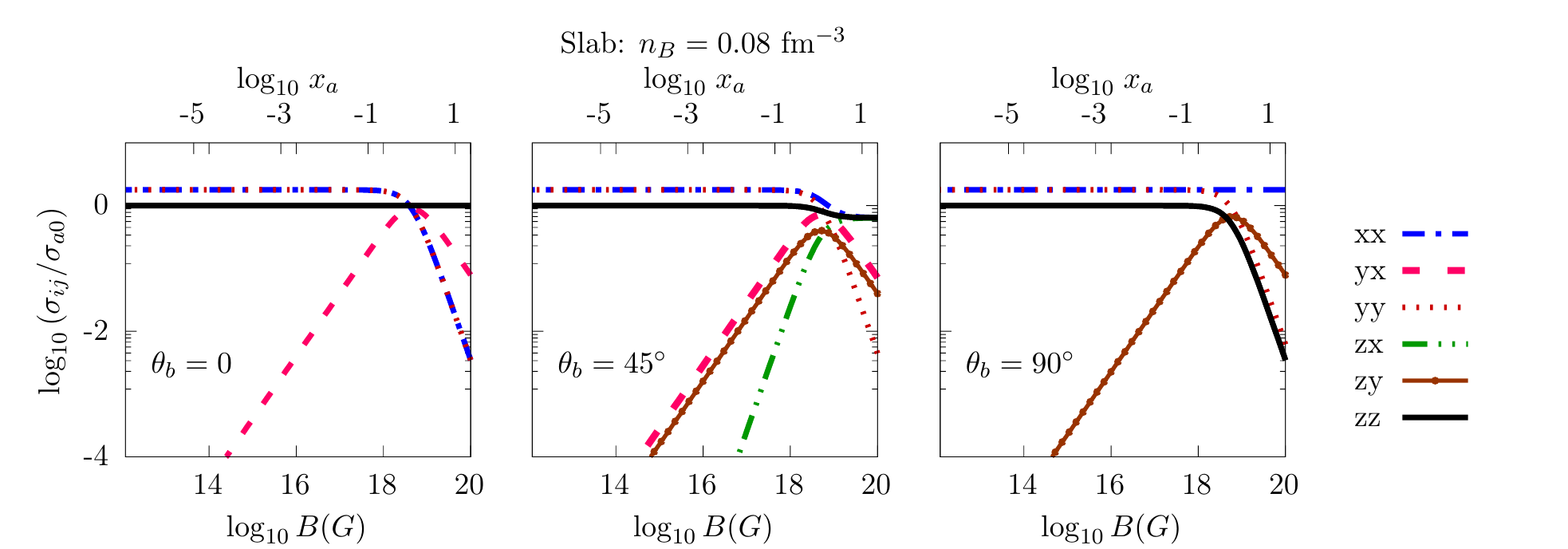}
    \caption{Components of the electron conductivity in units of the axial conductivity at $B=0$, as a function of the magnetic field for slabs at $n_B=0.08$ fm$^{-3}$.   The angle between the pasta symmetry axis and the magnetic field is fixed at 0 (left), 45º (center) and 90º (right). In the top axis, we show the variable $x_a = \omega/\nu_a$.}
    \label{fig:sigma_x1}
\end{figure*}

{We now turn to the effect of the magnetic field on the conductivities.
 When including the magnetic field, we show all components of the conductivity in units of the dominant conductivity at zero magnetic field, i.e. the perpendicular (axial) conductivity for rods (slabs).  In Figs.~\ref{fig:sigma_x2} (\ref{fig:sigma_x1}) we show the conductivity components when the magnetic field forms an angle $\theta_b=0, 45$º, and $90$º with the symmetry axis of the pasta. Different off-diagonal components appear depending on the angle of the magnetic field: if it lies in the symmetry axis of the pasta, only the perpendicular $xy$ component is not zero, the $zz$ component depends only on $\nu_a$, and the perpendicular $xx$ and $yy$ components are determined by both $\nu_p$ and $\bm{B}$. If it lies perpendicularly to the symmetry axis, only $zy$ is not zero, the $zz$ component is determined by $\nu_a$ and $\bm{B}$ and $xx=yy$ only by $\nu_p$. At  magnetic fields $B < 10^{18}$ G, the $zz$ component is not drastically modified, transverse components increase (decrease) for rods (slabs), and the off-diagonal terms increase steadily. At $\sim 10^{18}$ G, the diagonal components parallel to the magnetic field are unaffected, but the perpendicular and off-diagonal components start to decrease.} A magnetic field of this order is not far from the one expected at the very bottom of the inner crust of magnetars, which is about $20\%$  of the field in the core~\citep{PhysRevC.99.055811,Fujisawa2014}.

 For the average conductivity, we follow \citet{Yakovlev2016} once more, and assume the pasta takes random orientations with respect to the magnetic field since up to date there is no information regarding its orientation or prevalence of domains. To calculate the average parallel, perpendicular, and Hall terms we define a plane orthogonal to the magnetic field with the vectors $\bm{e}_1$, $\bm{e}_2 =  \bm{e}_1 \times \bm{b}$ and make the projections: $\sigma_\perp = \bm{b} \cdot \hat{\sigma} \cdot \bm{b}$, $\sigma_\| = \bm{e}_1 \cdot \hat{\sigma} \cdot \bm{e}_1$ and $\sigma_H = \bm{e}_1 \cdot \hat{\sigma} \cdot \bm{e}_2$, such that averaging the coefficients over all directions leads to:
 \begin{widetext}
\begin{equation}
    \begin{pmatrix}
    \langle \sigma_\perp \rangle \\
    \langle \sigma_\| \rangle \\
    \langle \sigma_H \rangle 
    \end{pmatrix}
    =  \frac{e^2 n_e}{m_e^\ast \omega^2}
    \begin{pmatrix}
    (\omega^2 +\nu_p \nu_a) (\nu_p^2 + \omega^2) H - \nu_p \\
    \frac{1}{2} [\nu_a \nu_p(\omega^2 - \nu_p^2)H + \nu_p] \\
    \omega ( 1- \nu_a \nu_p H)
    \end{pmatrix}
\text{ with }
    H =  \begin{cases}
     (sr)^{-1} \arctan(s/r)             & \nu_a>\nu_p \\
     (sr)^{-1} \arctanh(s/r)            & \nu_a < \nu_p \\
     (\nu_a^3 + \omega^2 \nu_a)^{-1}    & \nu_a = \nu_p
    \end{cases}
\end{equation}
\end{widetext}
$s = \omega \sqrt{|\nu_a - \nu_p|}$ and $r= \sqrt{\nu_p (\omega^2 + \nu_a \nu_p)}$. For $B \rightarrow 0$ we get 
\begin{equation}
    \langle \sigma_\perp \rangle=  \langle \sigma_\| \rangle = \frac{e^2 n_e}{m_e^\ast} \langle \nu^{-1} \rangle, \quad \langle \nu^{-1} \rangle = \frac{1}{3} \left( \frac{2}{\nu_p} + \frac{1}{\nu_a} \right)
\end{equation}
and the Hall parameters is zero $\langle \sigma_H \rangle \rightarrow 0$. One must notice that the average conductivity is proportional to the average of the inverse of $\langle \nu \rangle$,  and not $ \langle \nu \rangle$ itself, therefore its calculation does not amount to averaging the matrix element over $\Omega_q$.
\begin{figure*}
    \centering
    \includegraphics[scale=0.9]{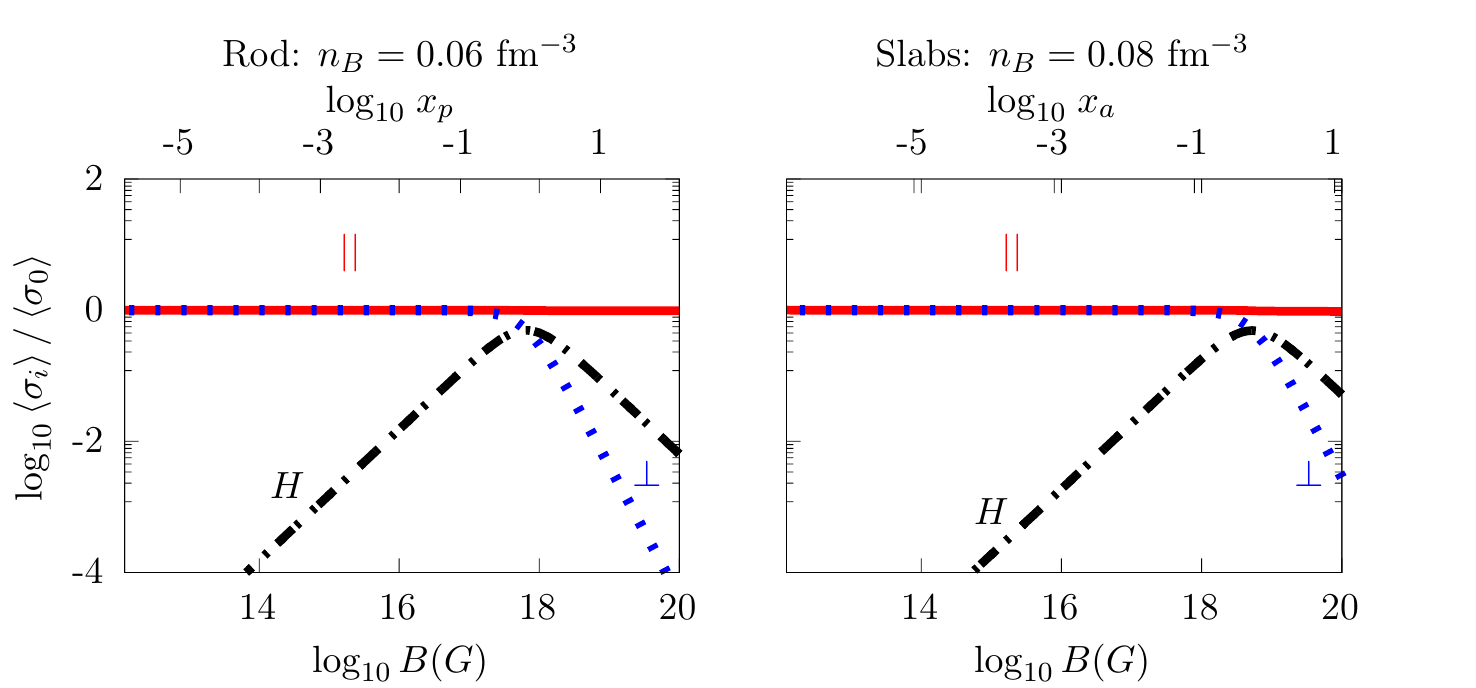}
      \caption{Average parallel, perpendicular and Hall conductivities for randomly oriented rods (left) and slabs (right) at $n_B=0.06$ fm$^{-3}$ and $n_B=0.08$ fm$^{-3}$, respectively. In the top axis, we show the variable $x_i = \omega/\nu_i$, with $i=p\; (a)$ on the left (right).}
    \label{fig:sigmaavg}
\end{figure*}

In Fig.~\ref{fig:sigmaavg} we show the average conductivities for rods and slabs, respectively, in units of the average conductivity with $B=0$. This figure can be compared with Fig.3 of~\citet{Yakovlev2016}, though the parallel component in our calculation is not so different from the perpendicular one due to the small difference we found between the anisotropic frequencies, unlike the assumption of~\citet{Yakovlev2016}.
To conclude this discussion, it is important to note that for all the calculations reported in this paper, the inner crust structure was computed without accounting for the magnetic field. Numerous studies exist in the literature addressing this point, using CLD or Thomas-Fermi techniques with different nuclear models, see e.g. \citet{Nandi2011,Lima_2013,Bao_2021,Wang_2022}. 
The general result of these works is that only extreme magnetic fields of the order of $B=10^{18}$ G affect the density profiles of the Wigner-Seitz cells, with an increased average proton fraction, particularly in the outer part of the inner crust dominated by spherical nuclei, and an increase of the charge of the pasta structures, that however does not exceed $\approx 10-20 \%$. These modifications would not affect the results presented in Figs. \ref{fig:sigma_x1} and ~\ref{fig:sigmaavg}, and would lead to an extra decrease of the conductivity in Fig.~\ref{fig:sigma_nb}, since $\sigma\propto Z^{-2}$, see eq.\eqref{eq:nuap}. 

\section{Conclusions}
\label{sec:conclusions}

In this paper, we studied the collision frequencies for elastic scattering between electrons and two different pasta phase structures. To do so, we performed an expansion of the collision integral in spherical harmonics, which allowed us to treat also the scattering with non-spherical targets. We applied this framework to calculate the electrical conductivity tensor.

The form factor of the pasta structures was evaluated by direct integration, although we neglected contributions from the structure factor,  which is equivalent to assuming that electron scatterings with different pasta targets are completely incoherent. This is a reasonable first approximation at high temperatures of the order of the MeV or above. More work is needed to evaluate the anisotropy of the transport coefficients at lower temperatures where the lattice long-range order along the pasta symmetry axis is likely to be preserved.

{
We find that anisotropic collision frequencies are highly dependent on the length of the pasta structures. In the high-temperature regime,  where the effective length that participates in the scattering is comparable to the WS length, the anisotropy is small and affects mainly the components of the conductivity perpendicular to the pasta symmetry axis.
}
It should be emphasized that neutron star properties are expected to be significantly impacted by the presence of different and mixed pasta geometries~\citep{2021PhRvC.103e5810C, PhysRevC.93.065806, newton_glassy_pasta}, and their (possibly disordered) mesoscopic arrangement.
Unfortunately, information is still lacking on how pasta domains, defects, and impurities appear at larger scales, but the presence of this kind of disorder is considered to be a likely feature of the pasta layers~\citep{2021PhRvC.103e5810C, PhysRevC.93.065806, newton_glassy_pasta, PhysRevC.104.L022801}. 
Our treatment, at the moment, does not include precise corrections coming from scattering with domain boundaries and mixed geometries. 
Future investigations of these matters must be incorporated within the present framework to improve it. 

Our numerical results are based on the IUFSU force, a simplified modelling of the pasta phase using a one-component liquid drop approach, and an estimation of the pasta sizes based on the values of the correlation between neighbouring WS cells.
However, the analytical results are general and can be used to calculate the transport properties of the inner crust, with and without a magnetic field, by using any microscopic estimation of the pasta linear dimensions and proton number from microscopic mean-field or molecular dynamics simulations.

\section*{Acknowledgements}

This  work is a part of the project INCT-FNA Proc. No. 464898/2014-5, and of the Master project In2p3 NewMAC. D.P.M. is  partially supported by Conselho Nacional de Desenvolvimento Cient\'ifico  e  Tecnol\'ogico  (CNPq/Brazil) under grant  303490/2021-7 and  M.R.P. is supported by Conselho Nacional de Desenvolvimento Científico e Tecnológico - Brasil  (CNPq)  and with a  doctorate  scholarship by Coordena\c c\~ao de Aperfei\c coamento de Pessoal de N\'ivel Superior (Capes/Brazil). M. R. P. also acknowledges partial support from LPC Caen.
MA acknowledges partial support from PHAROS, COST Action CA16214.  

\section{Data Availability} 
No new data were generated or analysed in support of this research.


\bibliographystyle{mnras}
\bibliography{refs}


\appendix

\section{Isotropic limit}
\label{app:iso_limit}

In this appendix, we show that the isotropic limit is obtained from eq.~\eqref{eq:nu_final} when $W_{pp'}$ is a function only of $|\bm{q}|$ 
The equation obtained is equivalent to to eq.~(3.135) of \citet{pines2018theory}.
In the isotropic case, there is no change in the $l$ index of spherical harmonics during the collision, and the sum of $m$ and $m'$ indexes are zero since $W_{pp'}$ is a function of the relative angle between $\bm{p}$ and $\bm{p'}$, therefore:
\begin{equation}\label{eq:W_iso}
\W_{lm\,l'm'} = \W_{l m} \delta_{ll'} \delta_{m\,-m'}.
\end{equation}
This can be understood from the expansion in eq.~\eqref{eq:Wexpansion}, where in the isotropic case the pair of spherical harmonics must be replaced by the Legendre polynomial. Eq.~\eqref{eq:nu_final} can be rewritten as
\begin{flalign}
& \left[\nu\right]_{lm}^{l'm'} =     \frac{p^2 }{4 \pi^2 v } \Bigg[(-1)^{m'} \sqrt{(2 l +1)(2l'+1)}\W_{00 \;00}
\begin{pmatrix}
l & l' &0 \\
0 & 0 & 0
\end{pmatrix} \nonumber \\
&\times \begin{pmatrix}
0 & l & l'\\
0 & m & -m'
\end{pmatrix}
-  \; (-1)^m \, \W_{l m\,l-m} \delta_{ll'} \delta_{m m'}\Bigg].
\end{flalign}
We simplify this expression by utilizing the following property of the 3j-Wigner symbols:
\begin{equation}
    \begin{pmatrix}
l & l' &0 \\
0 & 0 & 0
\end{pmatrix} \begin{pmatrix}
0 & l & l'\\
0 & m & -m'
\end{pmatrix}
= \frac{(-1)^{-2l +m} }{2l+1} \delta_{ll'}\delta_{m m'}
\end{equation}
and defining  $ \W_{lm}= \W_{l m\; l-m}$, such that:
\begin{equation}\label{eq:nutemp}
\left[\nu\right]_{lm}^{l'm'}=   \frac{p^2 }{4 \pi^2 v } \delta_{ll'}\delta_{m m'} \Bigg[\W_{00} - (-1)^m  \W_{l m} \Bigg].
\end{equation}

To recover the usual integral equation with the transition matrix element we use eq.~\eqref{eq:W_label}, with the aid of eq.~\eqref{eq:change_var}
\begin{equation}
\W_{lm} =  \frac{2 \pi}{p^2} \int  \frac{d^3\bm{q}}{q} W_{pp'}(-1)^{m} {{Y}_l^m} ^\ast (\Omega_p)  {{Y}_{l}^{m}}^\ast (\Omega_{p'})
\end{equation}
into eq.~\eqref{eq:nutemp} and average over the $m$ index:
\begin{equation}
\nu_l =  \frac{p^2 }{4 \pi^2 v (2l+1)} \int  \frac{d^3\bm{q}}{q}  W_{pp'}\sum_m \Bigg[\frac{1}{4 \pi} -   {{Y}_l^m} ^\ast (\Omega_p)  {{Y}_{l}^{m}}^\ast (\Omega_{p'} ) \Bigg].
\end{equation}
Using the property of spherical harmonics
\begin{equation}
    \sum_m {{Y}_l^m} ^\ast (\Omega_p)  {{Y}_{l}^{m}}^\ast (\Omega_{p'} ) = P_l(cos\xi) 
\end{equation} 
 where $\cos \xi = \bm{p}\cdot \bm{p'}/(|\bm{p}| |\bm{p'}|)$ and changing variables as $2 q dq = - p^2 d(\cos \xi)$, we obtain
\begin{equation}
\nu_l = \frac{p^2 }{4 \pi v } \int_{-1}^1 d(\cos \xi) W_{pp'} \Bigg[1  -   P_l (\cos \xi) \Bigg],
\end{equation}
which is equivalent to eq.~(3.135) of \citet{pines2018theory} for electron scattering with isotropic targets.  We use  $q^2 = p_F^2 (1-\cos \xi)$ to change variables and write, for $l=1$:
\begin{equation}
\nu_1(\upepsilon_p)= \frac{1}{4 \pi p^2 v } \int _0^{2 p} q^3 \, d q W_{pp'}, 
\end{equation}
Using eq.~\eqref{eq:W_scattering}, we recover eq.~\eqref{eq:nu_iso}. Likewise, the viscosity can be obtained with the $l=2$ --  see  eqs~(2) and (3) of ~\citet{Chugunov:2005nc}: 
\begin{equation}
\nu_2(\upepsilon_p)= \frac{3}{4 \pi p^2 v } \int _0^{2 p} q^3 \, d q \Big( 1 - \frac{q^2}{4p^2} \Big)W_{pp'}. 
\end{equation}

\label{lastpage}
\end{document}